\documentclass[12pt, draftclsnofoot, onecolumn]{IEEEtran}
\usepackage{graphicx}
\usepackage{color}
\usepackage{fancyhdr}
\usepackage[cmex10]{amsmath}
\usepackage{mdwmath}
\usepackage{amssymb}
\usepackage{textcomp}
\usepackage{times}
\usepackage{multicol}
\usepackage{float}
\usepackage[numbers,sort&compress]{natbib}
\usepackage{color, soul}
\usepackage{subfig}

\usepackage{color}
\usepackage{algorithm} 
\usepackage{algorithmic} 
\usepackage{multirow}
\ifCLASSINFOpdf
\else
\fi
\begin{document}
%
\title{Optimum Transmission Policies for Energy Harvesting Sensor Networks Powered By a Mobile Control Center}
%



\author{Tao~Li,~Pingyi~Fan,~\IEEEmembership{Senior~Member,~IEEE},~Zhengchuan~Chen, and~Khaled~Ben~Letaief,~\IEEEmembership{Fellow,~IEEE}
\thanks{This paper was presented in part at IEEE International Conference on Communication (ICC), June,
2015, London, UK.

T. Li, P. Y. Fan and Z. C. Chen are with Tsinghua National Laboratory for Information Science and Technology (TNList), and Department of Electrical Engineering, Tsinghua University, Beijing, P. R. China, 100084 (e-mail: litao12@mails.tsinghua.edu.cn; fpy@tsinghua.edu.cn; chenzc10@mails.tsinghua.edu.cn).

K. B. Letaief is with the Department of Electronic and Computer Engineering, Hong Kong University of Science and Technology, Sai Kung, Hongkong (e-mail: eekhaled@ece.ust.hk).}}

\maketitle
\thispagestyle{empty}
\baselineskip 24pt
\begin{abstract}
\baselineskip 18pt
Wireless energy transfer, namely RF-based energy harvesting, is a potential way to prolong the lifetime of energy-constrained devices, especially in wireless sensor networks. However, due to huge propagation attenuation, its energy efficiency is regarded as the biggest bottleneck to widely applications. It is critical to find appropriate transmission policies to improve the global energy efficiency in this kind of systems.
To this end, this paper focuses on the sensor networks scenario, where a mobile control center powers the sensors by RF signal and also collects information from them.
Two related schemes, called as \emph{harvest-and-use} scheme and \emph{harvest-store-use} scheme, are investigated, respectively. In \emph{harvest-and-use} scheme, as a benchmark, both constant and adaptive transmission modes from sensors are discussed. To \emph{harvest-store-use} scheme, we propose a new concept, the best opportunity for wireless energy transfer, and use it to derive an explicit closed-form expression of optimal transmission policy. It is shown by simulation that a considerable improvement in terms of energy efficiency can be obtained with the help of the transmission policies developed in this paper. Furthermore, the transmission policies is also discussed under the constraint of fixed information rate. The minimal required power, the performance loss from the new constraint as well as the effect of fading are then presented.
\end{abstract}


%
\IEEEpeerreviewmaketitle

\section{Introduction}
%
%
%
%

Traditional wireless sensor networks are usually constrained by limited battery energy, which is widely regarded as a fundamental performance bottleneck. Energy harvesting, as a promising solution to prolong system's lifetime, has drawn great attention recently, especially the energy harvesting technique based on radio frequency (RF) signal, which can provide a flexible and reliable energy flow without any intermittence. Since the idea of wireless energy transfer was brought up again \cite{Kurs_1}, there has been a rapidly growing interest in this field, e.g., \cite{Grover_2,Liu_3,Fan_4,Chen_5,Li_6,Ju_7}. In \cite{Grover_2}, simultaneous information and energy transfer was discussed in frequency selective channel. A novel dynamic receiver structure was proposed in \cite{Liu_3}. In \cite{Fan_4}, it considered the wireless relay energy transfer and information transmission in MIMO-OFDM systems. In \cite{Chen_5}, the energy beam-forming in multi-antenna system with limited feedback was explored. In \cite{Li_6}, the maximal achievable rate region for two-way information rates in a wireless powered sensor network was investigated. Besides, the work in \cite{Ju_7} studied the maximum uplink throughput when multiple terminals are powered by the control center in the downlink.

Most of these existing works focus on the static networks. However, many practical systems are working under a mobile scenario \cite{Shi_8,He_9}. For instance, there are a large number of sensor nodes deployed along high speed railway. A moving train passing by the sensors can power them by RF signal and read the monitoring information of system from them. As for the highway, the moving car can read the local information from the sensors on the roadside, which is wirelessly powered by the moving car. Another typical example is that an unmanned aerial vehicle travels through an expanse area that is deployed with energy harvesting sensor networks, collecting the data for geological and biological investigations.
Generally speaking, in wireless sensor networks powered by RF-based energy harvesting under a mobile scenario, propagation loss is much more serious than that in traditional systems \cite{Ju_7}, which may lead to a very poor energy efficiency. Thus, it is essential to employ appropriate strategies in mobile scenarios, namely to utilize the mobility and position information of terminals to enhance the energy efficiency of wirelessly powered system, which has not been considered in the literature to the best of our knowledge.

Based on the considerations above, this paper concentrates on the optimum transmission policy for RF-based energy harvesting sensor networks that are triggered off by a mobile control center, where the mobile control center collects the information generated/stored at each sensor node and the sensors are powered by the RF signal transmitted from control center without any other energy sources.
Two related schemes, called as \emph{harvest-and-use} scheme and \emph{harvest-store-use} scheme, are investigated in this paper, respectively. In \emph{harvest-and-use} scheme, where the energy needs to be consumed immediately after it has been harvested, both constant and adaptive transmission modes are discussed. In \emph{harvest-store-use} scheme, with the help of energy storage device at sensor node, such as a rechargeable battery, the harvested energy can be stored temporarily and then reused later. To express easily, we propose a new concept, the best opportunity of wireless energy transfer and use it as the key tool to derive an explicit closed-form expression of optimum transmission policy. Simulation results indicate that a considerable improvement can be observed by employing the transmission policy developed in this paper. Due to low complexity of implementation, the performance under fixed information rate constraint from sensors is also investigated carefully. The corresponding transmission policy is discussed in both deterministic and random channel models.

In fact, similar technologies have already been used in practical system, such as radio frequency identification devices (RFID) \cite{Bose_10,Liu_11}. However, since RFID system is usually quasi-static and transmission distance is relatively small, energy efficiency is not so important as the mobile scenario concerned in this paper. Another typical example is Radar system, which can monitor a target by detecting the energy of reflected wave that is transmitted by itself \cite{Yang_12}. It indicates that economic consideration (namely energy efficiency) rather than feasibility is the main obstacle for applying RF-based energy harvesting into practical system, which is the main motivation of this paper.
Besides, it is worth noting that the energy transfer and information feedback are jointly considered in this paper, which is different from some other optimum strategies in prior works, e.g., \cite{Sharma_13,Yang_14,Ding_15,Ho_16,Tutuncuoglu_17,Ozel_18}. It is also the reason that great improvement can be achieved by the policies developed in this paper. The main contributions of this paper are summarized as follows:

\begin{itemize}
\item{In harvest-and-use scheme, the energy is consumed immediately by the sensor node after it has been harvested. Both constant and optimal adaptive transmission modes are presented in terms of maximizing cumulative throughput.}
\item{In harvest-store-use scheme, wireless information and energy transfer are jointly optimized under the help of energy storage. An explicit closed-form expression of optimal transmission policy is established by using the concept of the best opportunity of wireless energy transfer.}
\item{For low implementation complexity, the transmission policies under fixed data rate constraint from sensors are studied in both deterministic and random fading channel models. The performance of them are discussed via simulation results.}
\end{itemize}

The rest of this paper is organized as follows. System structure and channel model are introduced in Section II. The optimal transmission policies in harvest-and-use scheme and harvest-store-use scheme are explored in Section III and Section IV, respectively. When data rate is constrained to a constant value, the corresponding strategy is discussed in Section V. Lastly, some simulation results and conclusions are given in Sections VI and VII, respectively. Besides, to make the formulation and discussion in the sequel more clear, the descriptions of all the notations used in this paper are given in TABLE \ref{table of parameters}.

\begin{table}
\centering
\caption{The descriptions of all the notations used in this paper.}\label{table of parameters}
\begin{tabular}{|c|c|}
\hline
PARAMETER  &  DESCRIPTIONS OF THE NOTATIONS\\
\hline
$G_c$, $G_s$ & Constant power gain in the wireless links  \\
\hline
$\alpha_c$, $\alpha_s$ & Propagation attenuation exponent in the wireless links \\
\hline
$p_c(t)$, $p_s(t)$ & Transmit power in the wireless links \\
\hline
$h_c(t)$, $h_s(t)$ & The coefficient of wireless channel fading \\
\hline
$p_v(t)$ & The equivalent transmit power at the virtual energy transmitter \\
\hline
$\sigma_0^2$ & The power value of additive channel noise \\
\hline
$\eta$ & The performance loss ratio resulted from maximum allowable power constraint \\
\hline
$\beta(t)$ & The random small-scale fading coefficient \\
\hline
$m$ & The parameter for Nakagami fading distribution \\
\hline
$\alpha_c$, $\alpha_s$& Path loss exponent in the wireless link from control center to sensor node \\
\hline
$P_{cons}$ & Baseband circuit power consumption at the node with energy harvesting\\
\hline
$v_0$ & The value of the velocity of moving control center \\
\hline
$\xi$ & The coefficient of energy transformation efficiency at the energy harvesting node\\
\hline
$d_m$ & The radius of effective coverage of each sensor $|OA|=|OC|$ in Fig. \ref{fig:system_structure for unit}  \\
\hline
$d_0$ & The distance between sensor and moving track $|OB|$ in Fig. \ref{fig:system_structure for unit}  \\
\hline
$L_0$ & The moving range $|AB|=|BC|$ in Fig. \ref{fig:system_structure for unit} \\
\hline
$R$ & The number of data that can be transmitted from sensor to the control center during a whole period  \\
\hline
\end{tabular}
\end{table}

\section{Preliminary}

\subsection{System Structure and Parametrization}

Fig. 1(a) illustrates the diagram of system structure, which consists of a mobile control center and plenty of sensor nodes. The sensors are wirelessly powered by the control center without any other energy sources. And the control center is moving through the area of interest along a line with constant speed $v_0$, to collect the data from sensor nodes.
Each sensor node has an effective coverage due to the sensitivity of transceiver, denoted as dashed circle in Fig. 1(a). Within it, the control center can establish a physical link for transmission. It is assumed that the coverage of different sensor node does not overlap with each other so that there is no interference between them. The results derived here can be straightforwardly extended to overlap-coverage case by some orthogonal multiplexing division techniques for the sensor nodes within the overlapped coverage.
Based on the assumptions above, we can focus on the data collecting process of particular sensor node, which can capture all the insights of the system.

To express it easily, Fig. 1(b) shows a simplified unit for the system, in which the mobile control center is under the coverage of a specific sensor node.
A two-dimension model can be adopted to formulate the problem instead of the original three-dimension model in Fig. 1(a), since the sensor node and moving track of the control center can be involved into a plane. In Fig. 1(b), $ABC$ denotes the track of the control center. Let system time $t$ be $0$ when the control center is passing the point $B$. The distance between the sensor and moving track is $|OB|=d_0$. The coverage radius is $|OA|=|OC|=d_m$ and $d_m=\sqrt{d_0^2+L_0^2}$ (where $|AB|=|BC|=L_0$). Then, the distance between control center and the sensor at system time $t$ is
\begin{equation}\label{eqn:transmission distance}
  d(t)=\sqrt{d_0^2+(v_0t)^2},\,\,\,t\in [-\tfrac{L_0}{v_0},\tfrac{L_0}{v_0}].
\end{equation}

%

\begin{figure}[!t]
\centering
\subfloat[]{
\label{fig:system_structure for whole}
\begin{minipage}[t]{0.5\textwidth}
\centering
\includegraphics[width=2.5 in]{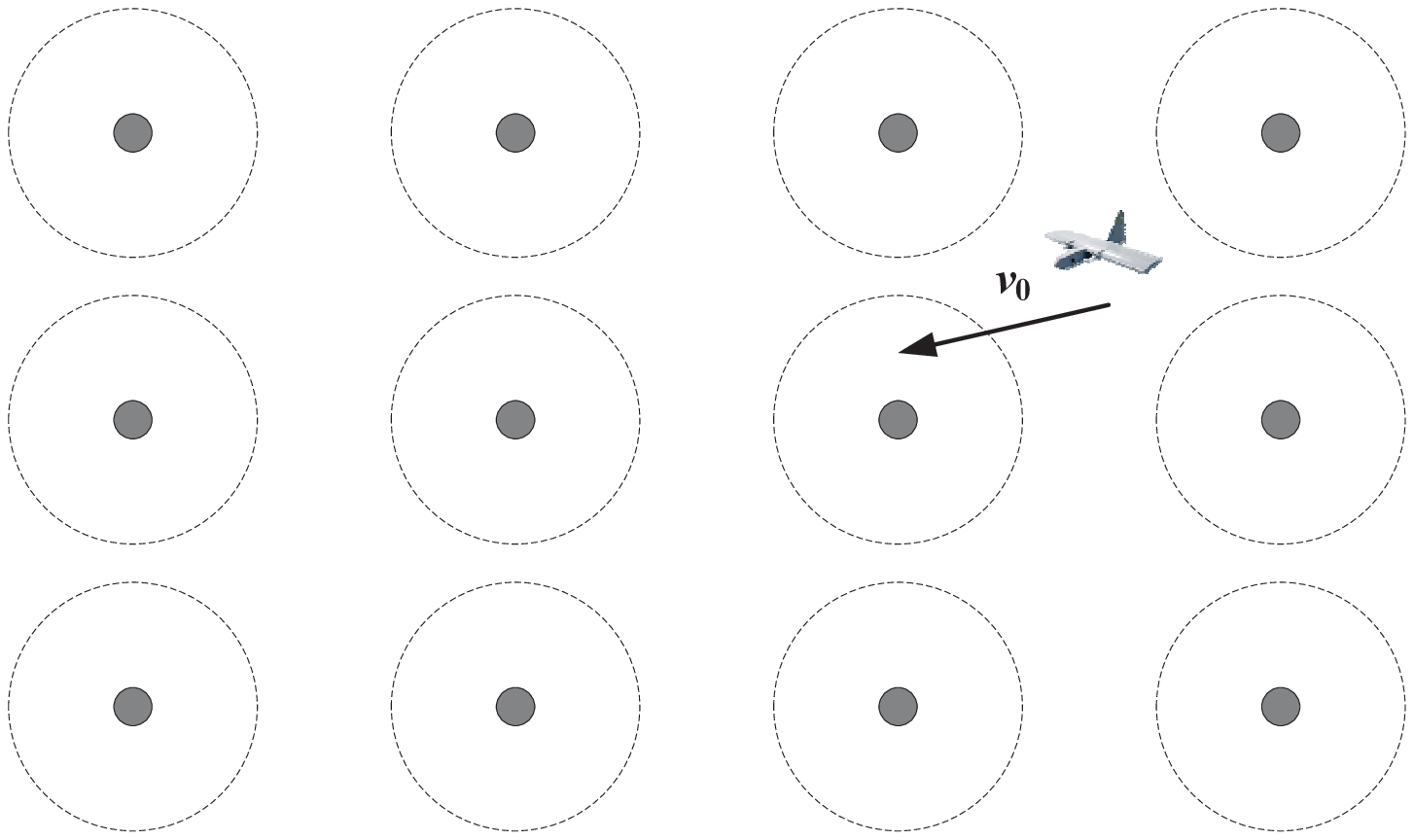}
\end{minipage}
}
\subfloat[]{
\label{fig:system_structure for unit}
\begin{minipage}[t]{0.5\textwidth}
\centering
\includegraphics[width=2.5 in]{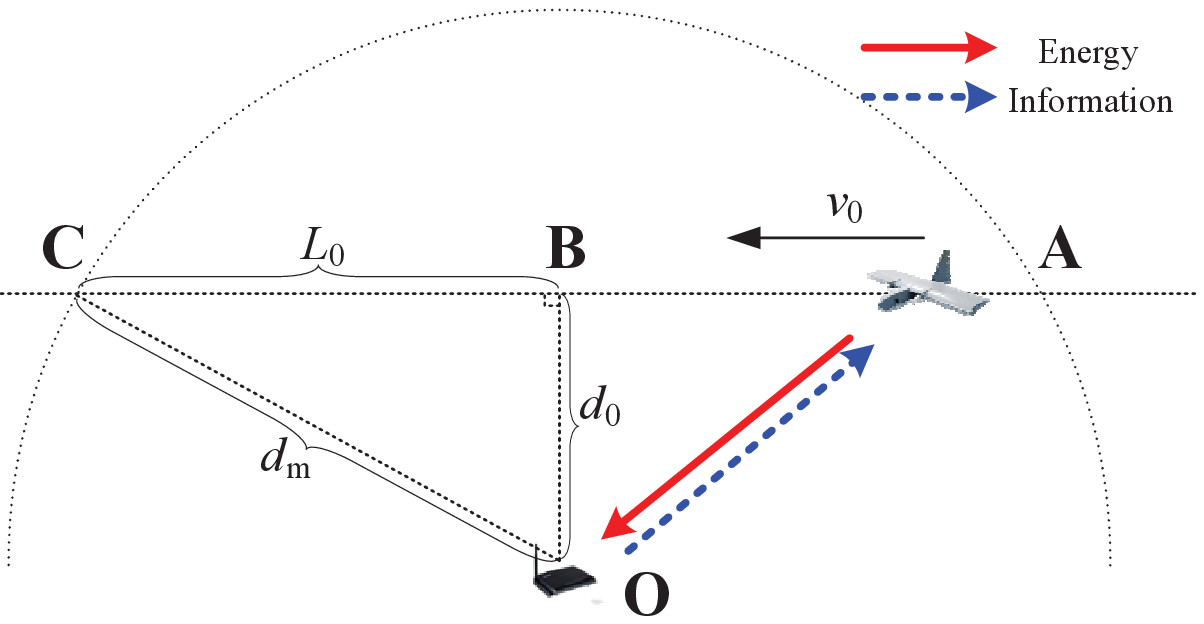}
\end{minipage}
}
\caption{The diagram of system structure for RF-based energy harvesting sensor networks: (a) An overall view of the network, (b) a simplified unit system with one sensor node.}
\end{figure}

\subsection{Channel Model for Energy Transfer and Information Transmission}

For wireless energy transfer from control center to the sensor, let $x_c(t)$ be the transmit signal with unit effective power and $p_c(t)$ be the transmit power at the control center. Through propagation attenuation, the received signal at system time $t$ can be modeled as \cite{Huang_19,Tse_20}
\begin{equation}\label{eqn:received signal at sensor}
  y_c(t)=\sqrt{ p_c(t)} h_c(t) x_c(t)+n_c(t), \,\,t\in [-\tfrac{L_0}{v_0},\tfrac{L_0}{v_0}],
\end{equation}
where $h_c(t)$ and $n_c(t)$ represent channel fading coefficient and channel noise, respectively.

Based on energy conservation principle, the energy profile that can be harvested from received signal at sensor node is \cite{Zhou_21} (where $\mathbb{E}[\cdot]$ denotes statistical mean operation)
\begin{equation}\label{eqn:harvested energy at sensor node}
  p_h(t)=\xi \mathbb{E}[|y_c(t)|^2]= \xi |h_c(t)|^2 p_c(t),\,\,t\in [-\tfrac{L_0}{v_0},\tfrac{L_0}{v_0}],
\end{equation}
where $\xi$ is the coefficient of energy harvesting efficiency at sensor node, the value of which is regarded as $50\%$ without any other declaration in this paper. The power contributed by additive noise is ignored due to Thermodynamic Law limit \cite{Zhou_21}.

For information transmission from sensor node to control center, the channel model is
\begin{equation}\label{eqn:information channel model}
  y_s(t)=\sqrt{p_s(t)} h_s(t) x_s(t)+n_s(t), \,\,t\in [-\tfrac{L_0}{v_0},\tfrac{L_0}{v_0}],
\end{equation}
where $x_s(t)$ and $p_s(t)$ are transmit signal and transmit power from sensor node at system time $t$, ($\mathbb{E}[|x_s(t)|^2]=1$). $n_s(t)$ is additive white Gaussian noise with zero mean and $\sigma_0^2$ variance.

\subsection{Some Basic Assumptions in This Paper}

For analysis tractability, several basic assumptions are adopted in this paper, which are listed as follows.

\emph{Assumption 2.1: The control center is in uniform linear motion with speed $v_0$, and the relative position between control center and sensor is precisely known by control center. Besides, the time and frequency synchronization between control center and sensor nodes is ideal.}

Generally speaking, the control center usually has a map about the accurate position information of all the affiliated sensors. Besides, many advanced techniques, such as Global Position System (GPS), accelerometer, goniometer et al., can help the control center to determine its position. Thus, we assume that the relative position information, namely $d(t)$, is precisely known at control center. It needs to mention that the specific algorithm for positioning is beyond the scope of our work, and this paper just concentrates on how to utilize available position information to achieve a better energy efficiency in a mobile scenario.

\emph{Assumption 2.2: Signal transmitting and baseband circuit are two main energy consumptions at sensor nodes. It is assumed that the circuit power consumption, such as for sensing and data processing, is a constant value $P_{cons}$ when the sensor node is under the working status [22]. That is to say, $P_{cons}$ energy flow will be consumed by baseband circuit during the period that the sensor node is activated. Otherwise, sensor node keeps silent.}

Besides, it needs to mention that the system model concerned in this paper is a centralized architecture, in which the associated sensor node works under the control of the center node. Namely, once the sensor node has been woke up, it can receive the system information (or control command) continuously via the wireless channel from the mobile control center.

\section{Optimum Transmission Policy in Harvest-and-Use Scheme}

In harvest-and-use scheme, the energy is consumed immediately once it is harvested, (where we assume that the time-delay caused by physical circuit can be ignored). Otherwise, it will be wasted. This scheme is suitable for the system with no energy storage. Obviously, power allocation strategy at control center is the main problem need to explore in this condition. Both the constant and adaptive transmission policy will be presented in the sequel, respectively.

\subsection{Constant Transmission Policy in Harvest-and-Use Scheme}

As a baseline, constant transmission policy (CTP) at control center is presented firstly, which is simplest and widely used in traditional static systems. The control center transmits signal with invariant power $P_0$ during the whole period $t\in [-\tfrac{L_0}{v_0},\tfrac{L_0}{v_0}]$.
The instantaneous rate capacity (in bit/Hz/s) of the channel from sensor to control center shown in (\ref{eqn:information channel model}) at system time $t$ is
\begin{equation}\label{eqn:information rate in harvest-and-use}
  r(t)=\log_2 \Big(1+\frac{|h_{s}(t)|^2 p_{s}(t)}{\sigma_0^2}\Big),\,\,t\in \big[-\frac{L_0}{v_0},\frac{L_0}{v_0}\big].
\end{equation}

When signal transmission and circuit energy consumption at sensor node are taken into consideration, the constraint for $p_{s}(t)$ under the harvest-and-use scheme can be written as
\begin{equation}\label{eqn:relationship between two powers in harvest-and-use}
  p_s(t)+P_{cons} I_{\{p_s(t)>0\}}\leq p_h(t),\,t\in \big[-\frac{L_0}{v_0},\frac{L_0}{v_0}\big],
\end{equation}
where $I_{\{x\}}$ is indication function, the value of which is $1$ if $x$ is $true$, otherwise, it is $0$.

To compare the performance of different transmission policies, the concept of cumulative throughput is adopted as a metric and is defined as follows.

\emph{Definition 3.1: The cumulative throughput $R$ (in bit/Hz) is defined as the information that can be successfully transmitted from sensor to the control center during a whole period $t \in [-\frac{L_0}{v_0},\frac{L_0}{v_0}]$.}

Substituting (\ref{eqn:harvested energy at sensor node}) and (\ref{eqn:relationship between two powers in harvest-and-use}) into (\ref{eqn:information rate in harvest-and-use}), and $p_c(t)=P_0$, the cumulative throughput with constant transmission policy in harvest-and-use scheme is
\begin{equation}\label{eqn:throughput in harvest-and-use}
  R_{\textrm{\tiny CTP}}^{\textrm{\tiny HAU}}=\int_{-\frac{L_0}{v_0}}^{\frac{L_0}{v_0}} r(\tau) d\tau
  =\int_{-\frac{L_0}{v_0}}^{\frac{L_0}{v_0}} \log_2 \Big(1+\big(\frac{\xi |h_s(t)|^2 |h_c(t)|^2 P_0}{\sigma_0^2} - \frac{|h_s(t)|^2 P_{cons}}{\sigma_0^2} \big)^+\Big) d\tau,
\end{equation}
where the operation $(x)^+$ denotes the larger one between $x$ and $0$.

\subsection{Adaptive Transmission Strategy in Harvest-and-Use Scheme}

In a time-varying channel environment, adaptive power allocation strategy can achieve a better performance \cite{Goldsmith_23}. Let $p_c(t)$ be the transmit power of control center at system time $t$. For fairness, assuming the average value of $p_c(t)$ during $t\in [-\tfrac{L_0}{v_0},\tfrac{L_0}{v_0}]$ is still $P_0$. Under harvest-and-use scheme, the maximal cumulative throughput (OTP) can be formulated as
\begin{subequations}\label{equ:optimization problem for throughput maximization without battery}
\begin{align}
   R_{\textrm{\tiny OTP}}^{\textrm{\tiny HAU}}\,\,&= \,\max \limits_{p_c(t)} \Big\{ \int_{-\frac{L_0}{v_0}}^{\frac{L_0}{v_0}}\log_2 \Big(1+\frac{|h_s(t)|^2 p_s(\tau)}{\sigma_0^2}\Big)d\tau \Big\}   \tag{8}\\
   s.t.\,\,\,\,\,& \int_{-\frac{L_0}{v_0}}^{\frac{L_0}{v_0}} p_c(\tau)d\tau \leq P_0 \cdot \frac{2L_0}{v_0}, \label{equ:optimization problem for throughput maximization without battery a} \\
   & p_s(t)+P_{cons} I_{\{p_s(t)>0\}} \leq \xi |h_c(t)|^2 p_c(t),\,\,t\in [-\tfrac{L_0}{v_0}, \tfrac{L_0}{v_0}], \label{equ:optimization problem for throughput maximization without battery b} \\
   & p_s(t) \geq 0,p_c(t)\geq 0, \,t\in [-\tfrac{L_0}{v_0},\tfrac{L_0}{v_0}]. \label{equ:optimization problem for throughput maximization without battery c}
\end{align}
\end{subequations}
where the inequality in (\ref{equ:optimization problem for throughput maximization without battery a}) denotes average transmit power constraint at control center, the constraint in (\ref{equ:optimization problem for throughput maximization without battery b}) is the causality constraint for energy consumption at sensor node, and the inequality in (\ref{equ:optimization problem for throughput maximization without battery c}) reflects the nonnegativity property of transmit power.

\emph{Lemma 3.1: To maximize the energy efficiency, the transmit power at control center $p_c(t)$ at system time $t$ should be either more than $\frac{P_{cons}}{\xi |h_c(t)|^2}$ or equal to zero in harvest-and-use scheme.}
\begin{IEEEproof}
 See Appendix A.
\end{IEEEproof}

In this section and next section, a deterministic propagation model is adopted \cite{Tse_20}, in which $h_c(t)=\sqrt{\frac{G_c}{d(t)^{\alpha_c}}}$ and $h_s(t)=\sqrt{\frac{G_s}{d(t)^{\alpha_s}}}$, (where $G_c, G_s$ represent the constant antenna gain, and $\alpha_c, \alpha_s$ denote the propagation attenuation exponent, $2 \leq \alpha_c, \alpha_s \leq 5$).
Then, the optimal transmission strategy at control center in harvest-and-use scheme can be derived, which is

\emph{Proposition 3.1: The optimal adaptive strategy at control center in harvest-and-use scheme is }
\begin{equation}\label{eqn:optimal power allocation without battery}
  p_c^\ast(t)=\Big(\frac{1}{\lambda_1 \ln 2}-\frac{d(t)^{(\alpha_s+\alpha_c)}\sigma_0^2}{\xi G_s G_c} +\frac{d(t)^{\alpha_c} P_{cons}}{\xi G_c} \Big)^+,\,t\in \Big[-\frac{L_0}{v_0},\frac{L_0}{v_0}\Big],
\end{equation}
\emph{where the constant value $\lambda_1$ is determined by the constraint in (\ref{equ:optimization problem for throughput maximization without battery a}) with equality.}
\begin{IEEEproof}
  See Appendix B.
\end{IEEEproof}

Substituting (\ref{eqn:optimal power allocation without battery}) into (\ref{equ:optimization problem for throughput maximization without battery}), the cumulative throughput with optimal transmission policy in harvest-and-use scheme can be expressed as
\begin{equation}\label{eqn:total throughput without battery}
  R_{\textrm{\tiny OTP}}^{\textrm{\tiny HAU}} =\int_{-\frac{L_0}{v_0}}^{\frac{L_0}{v_0}} \log_2 \Big[ \max \Big(\frac{\xi G_s G_c}{\lambda_1 \ln 2 \cdot d(\tau)^{(\alpha_s+\alpha_c)}\sigma_0^2},1\Big) \Big] d\tau.
\end{equation}

\section{Optimum Transmission Policy in Harvest-store-use Scheme}

In harvest-store-use scheme, some harvested energy can be stored into buffer temporarily and then reutilized when the channel status is better. This scheme is suitable for the system that has an energy storage device at sensor node, such as a rechargeable battery \cite{Luo_24}. Better performance can be achieved by jointly considering the energy transfer and information transmission.
\subsection{Problem Formulation}

According to the state of art, it is reasonable to assume that the capacity of the battery is infinite compared with the energy that can be harvested during the considered period \cite{Ozel_25}. The causality constraint for energy consumption at sensor node in this case can be written as
\begin{equation}\label{eqn:the casuality constraint in harvest-store-use scheme}
  \int_{-\frac{L_0}{v_0}}^{t} p_s(\tau) d \tau +\int_{-\frac{L_0}{v_0}}^{t} P_{cons} I_{\{p_s(\tau)>0\}} d\tau \leq \int_{-\frac{L_0}{v_0}}^{t}p_h(\tau)d\tau,\,\,t\in \big[-\frac{L_0}{v_0},\frac{L_0}{v_0}\big].
\end{equation}

Let $p_c(t)$, $p_h(t)$ and $p_s(t)$ be instantaneous transmit power at control center, harvested energy and transmit power at sensor node, respectively. Provided that the average transmit power at control center is constrained to $P_0$, the maximal cumulative throughput with optimal transmission policy in harvest-store-use scheme can be formulated as the solution to the following problem
\begin{subequations}\label{equ:optimization problem for throughput maximization}
\begin{align}
   R_{\textrm{\tiny OTP}}^{\textrm{\tiny HSU}}\,&= \,\max \limits_{p_c(t),p_s(t)} \Big\{ \int_{-\frac{L_0}{v_0}}^{\frac{L_0}{v_0}}\log_2 \Big(1+\frac{|h_s(t)|^2 p_s(\tau)}{\sigma_0^2}\Big)d\tau \Big\}   \tag{12}\\
   s.t.\,\,\,& \int_{-\frac{L_0}{v_0}}^{t} \big( p_s(\tau)+P_{cons} I_{\{p_s(\tau)>0\}}\big) d\tau \leq \int_{-\frac{L_0}{v_0}}^{t}p_h(\tau)d\tau,\,t\in \big[-\frac{L_0}{v_0},\frac{L_0}{v_0}\big], \label{equ:optimization problem for throughput maximization a} \\
   & \int_{-\frac{L_0}{v_0}}^{\frac{L_0}{v_0}} p_c(\tau)d\tau \leq P_0 \cdot \frac{2L_0}{v_0}, \label{equ:optimization problem for throughput maximization c}\\
   & p_c(t),p_s(t) \geq 0 ,\,t\in [-\tfrac{L_0}{v_0},\tfrac{L_0}{v_0}].  \label{equ:optimization problem for throughput maximization d}
\end{align}
\end{subequations}
where the constraint in (\ref{equ:optimization problem for throughput maximization a}) denotes the causality relationship between harvested energy and consumed energy at sensor node, the constraint in (\ref{equ:optimization problem for throughput maximization c})
is average transmit power constraint at control center, and constraint in (\ref{equ:optimization problem for throughput maximization d}) reflects the nonnegativity of transmit power.

Recalling the system model in Fig. \ref{fig:system_structure for unit}, it needs to mention that an eligible transmission policy should not only clearly instruct the control center how to allocate the transmit power, but also instruct the sensor node how to manage the harvested energy and how to allocate its power for feedback transmission. Besides, since $p_c(t)$ and $p_h(t)$ have a deterministic relationship as given in (\ref{eqn:harvested energy at sensor node}), the problem in (\ref{equ:optimization problem for throughput maximization}) needs to be optimized over the range $\{p_c(t)\} \times \{p_s(t)\}$ that meet all the constraints in (\ref{equ:optimization problem for throughput maximization a}-\ref{equ:optimization problem for throughput maximization d}), which is a difficult problem to solve.

Note that the problem addressed above is different from some prior works in \cite{Sharma_13}--\cite{Ozel_18}, which mainly considered the information transmission problem powered by a random energy harvesting source. In particular, the work in \cite{Ho_16}--\cite{Ozel_18} derived the closed-form expression of optimal transmission policy, named as directional water-filling algorithm (which is named as staircase water-filling in \cite{Ho_16}), on the assumption that the profile of random energy is fully known at the beginning of transmission. However, the information transmission and wireless energy transfer are jointly considered in this paper. These results derived in \cite{Ho_16}--\cite{Ozel_18} can not be directly extended to the problem in (\ref{equ:optimization problem for throughput maximization}).
While preparing this manuscript, the authors became aware of a parallel work \cite{Zhou_26}, which also considered the optimal power allocation problem in a time-varying slot-based discrete system with the assumption that all the channel information is known by the system non-causally at the beginning of transmission.
The problem formulated in \cite{Zhou_26} is similar to one of the problems discussed in this paper, namely the optimization problem in (\ref{equ:optimization problem for throughput maximization}). However, this paper (the short version of this work is \cite{Li_27}) starts from a very different basic research problem, namely utilizing the mobility and position information of the terminal to enhance the energy efficiency of wireless powered information system. Apparently, the system model is time-continuous and we independently solve the problem (\ref{equ:optimization problem for throughput maximization}) by the virtual transmitter method that will be proposed in the next subsection. The authors think that both of these two works can provide some useful insights from different views for wireless powered system design.

\subsection{Some Properties of Harvested Energy Management}

Since the causality constraint for energy managing at the sensor node shown in (\ref{equ:optimization problem for throughput maximization a}) is the main obstacle to problem solving, we start from studying the properties of optimal energy managing strategy at the energy harvesting sensor node. Let $\Delta t$ be an arbitrary small time segment. Under the help of lossless energy buffer, the nature of a good energy management strategy is to store some energy $p_h(t_1)\Delta t$ harvested at system time $t_1$ into battery, and reuse it later in a more efficient time $t_2$ to obtain better system performance, (obviously $t_2\in[t_1, \frac{L_0}{v_0}]$). Since it is very difficult to find the optimal reuse time opportunity for the energy $p_h(t_1)\Delta t$ harvested at $t_1$, namely to obtain the closed-form expression of optimal value of $t_2$ as a function of $t_1$ on the condition that the value of $t_1$ is fixed, the generalized solution can not be explicitly expressed by using traditional way.

To do so, unlike traditional methods, we will consider this problem from a reverse way. The question can be described as that if the sensor node consumes $p_p(t_2) \Delta t$ energy at system time $t_2$, (which contains both transmit power consumption and circuit energy consumption), when is the best opportunity for this part of energy to be transferred from control center to sensor node? Namely, to obtain the expression of optimal value of $t_1$ on the condition that the value of $t_2$ is fixed. From a perspective of maximizing the whole energy efficiency, two interesting observations can be drawn, which are expressed as the following two Lemmas.

\emph{Lemma 4.1: When $t_0\in[-\frac{L_0}{v_0},0]$, the energy $p_p(t_0)\Delta t$ that is consumed by senor node at system time $t_0$ should be transferred from control center to sensor node at system time $t_0$. That is to say, the energy should be consumed at once it has been harvested during this period.}
\begin{IEEEproof}
 See the Appendix C.
\end{IEEEproof}


\emph{Lemma 4.2: When $t_0\in[0,\frac{L_0}{v_0}]$, the energy $p_p(t_0)\Delta t$ that is consumed by senor node at system time $t_0$ should be transferred from control center to sensor node at the system time $t=0$.}
\begin{IEEEproof}
 See the Appendix D.
\end{IEEEproof}

\begin{figure}[!t]
\centering
\includegraphics[width=2.8 in]{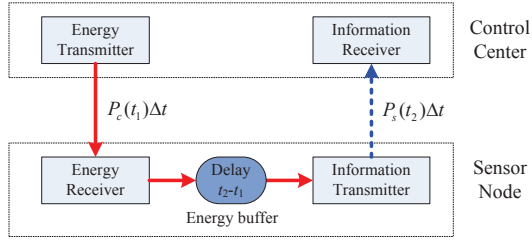}
\caption{The diagram of the wireless energy flow from control center to sensor node when there is an infinite energy storage at the sensor node.}
\label{fig:equivalent_model}
\end{figure}

In summary, the basic idea of Lemma 4.1 and Lemma 4.2 is to make full use of opportunistic wireless energy transfer to achieve the largest energy transfer efficiency, which reflects the optimal energy managing strategy at sensor node. Intuitively, the energy buffer acts as an energy delayer to avoid big propagation loss, which is illustrated in Fig. \ref{fig:equivalent_model}.

\subsection{Problem Solving}

To formulate the solution conveniently, we construct an auxiliary variable $p_v(t)$ to take place of $p_c(t)$. Specifically, it looks as if there is a virtual energy transmitter that transfers energy with power $p_v(t)$ at system time $t$, and the sensor node uses the energy harvested from $p_v(t)$ for information transmission immediately after it has been harvested.
In other words, we have artificially constructed an equivalent harvest-and-use system, after the energy management strategy has been decoupled from the initial problem with the help of virtual energy transmitter.
Provided that the sensor node consumes $p_p(t_2)\Delta t$ energy at system time $t_2$, the optimal opportunistic time of transferring this part of energy is $t_1$. Then the corresponding transmit power for virtual energy transmitter at $t_2$ is $p_v(t_2)=\frac{P_p(t_2)}{\xi |h_c(t_1)|^2}$ since the exact propagation attenuation for the energy $p_p(t_2)\Delta t$ is $|h_c(t_1)|^2$. Specifically, according to the results in Lemma 4.1 and Lemma 4.2 under the deterministic channel model, when $t_2\in[-\frac{L_0}{v_0},0]$, $P_v(t_2)=\frac{d(t_2)^{\alpha_c}}{\xi G_c}P_p(t_2)$. When $t_2\in[0,\frac{L_0}{v_0}]$, $P_v(t_2)$ is $P_v(t_2)=\frac{d_0^{\alpha_c}}{\xi G_c}P_p(t_2)$.

Of course, the average value of $p_v(t)$ during the concerned period should keep the same value as that of $p_c(t)$ due to the average power constraint at control center, which is
\begin{equation}\label{eqn:average value constraint for virtual transmit power}
  \int_{-\frac{L_0}{v_0}}^{\frac{L_0}{v_0}} p_v(\tau)d\tau = \int_{-\frac{L_0}{v_0}}^{\frac{L_0}{v_0}} p_c(\tau)d\tau. 
\end{equation}

Based on all the results developed above, we can establish a deterministic relationship between the optimal virtual transmit power $\{p_v^\ast(t)\}$ and optimum transmission policy for the system shown in Fig. \ref{fig:system_structure for unit}, which can be summarized as follows.

\emph{Proposition 4.1: Provided that $\{p_v^\ast(t)\}$ is the optimal allocation strategy for virtual energy transmitter, according to Lemma 4.1 and Lemma 4.2, the optimum transmission policy in harvest-store-use scheme in terms of cumulative throughput maximization during the whole period, namely the energy profile at control node $\{p_c^\ast(t)\}$ and the energy profile at sensor node $\{p_s^\ast(t)\}$, can be expressed as}
\begin{equation}\label{eqn:optimal power allocation at control center}
  p_c^\ast(t)=
\begin{cases}
   \,\,\,\,\,\,\,\,\,\,\,\,\,\,\,\,\,\,\,\,p_v^\ast(t),\,\,\,\,\,\,\,\,\,\,\,\,\,\,\, t\in [-\frac{L_0}{v_0},0) \\
   \int_0^{\frac{L_0}{v_0}} p_v^\ast(\tau)d\tau \cdot \delta(t), \,\,t=0 \\
   \,\,\,\,\,\,\,\,\,\,\,\,\,\,\,\,\,\,\,\,\,\,\,\,0,\,\,\,\,\,\,\,\,\,\,\,\,\,\,\,\,\,\,\,\,\, t\in (0,\frac{L_0}{v_0}]
\end{cases}
\end{equation}
\begin{equation}\label{eqn:optimal power allocation at sensor node}
  p_s^\ast(t)=
\begin{cases}
   \big(\frac{\xi G_c p_v^\ast(t)}{d(t)^{\alpha_c}}-P_{cons}\big)^+,\,\, t\in [-\frac{L_0}{v_0},0] \\
   \big(\frac{\xi G_c p_v^\ast(t)}{d_0^{\alpha_c}}-P_{cons}\big)^+,\,\, t\in [\,\,\,0\,\,,\frac{L_0}{v_0}]
\end{cases}
\end{equation}

The only residual task is to derive the expression of $p_v^\ast(t)$, since both $p_c^\ast(t)$ and $p_s^\ast(t)$ have deterministic relationships with $p_v^\ast(t)$. As stated previously, the causality constraint in (\ref{equ:optimization problem for throughput maximization a}) can be eliminated since the effect of optimal energy management strategy is reflected in $p_v(t)$. Thus, with the help of $p_v(t)$, the original problem in (\ref{equ:optimization problem for throughput maximization}) can be transformed as
\begin{subequations}\label{equ:the degradation of optimization problem for throughput maximization}
\begin{align}
   R_{\textrm{\tiny OTP}}^{\textrm{\tiny HSU}}\,\,\, &= \,\max \limits_{p_v(\tau)} \Big\{ \int_{-\frac{L_0}{v_0}}^{0} \log_2 \Big(1+\Big(\frac{\xi |h_s(\tau)|^2 |h_c(\tau)|^2 p_v(\tau)}{\sigma_0^2} - \frac{|h_s(\tau)|^2 P_{cons}}{\sigma_0^2} \Big)^+\Big)d\tau \tag*{}\\
   &\,\,\,\,\,\,\,\,\,\,\,\,\,\,\,\,\,\,+\int_{0}^{\frac{L_0}{v_0}}  \log_2 \Big(1+\Big(\frac{\xi |h_s(\tau)|^2 |h_c(0)|^2 p_v(\tau)}{ \sigma_0^2} - \frac{|h_s(\tau)|^2 P_{cons}}{ \sigma_0^2} \Big)^+ \Big) d\tau \Big\} \tag{16}\\
   s.t.\,\,\,\,\,& \int_{-\frac{L_0}{v_0}}^{\frac{L_0}{v_0}}p_v(\tau)d\tau \leq P_0 \cdot \frac{2L_0}{v_0}, \,\,\,\,\,\,\,\,\, \label{equ:the degradation of optimization problem for throughput maximization constraint a} \\
   & p_v(t)\geq 0,\,t\in \Big[-\frac{L_0}{v_0}, \frac{L_0}{v_0}\Big].  \label{equ:the degradation of optimization problem for throughput maximization constraint b}
\end{align}
\end{subequations}

Using similar discussion to the problem in (\ref{equ:optimization problem for throughput maximization without battery}), we firstly introduce a Lemma to isolate the effect of operation $(\cdot)^+$ in (\ref{equ:the degradation of optimization problem for throughput maximization}).

\emph{Lemma 4.3: To maximize the energy efficiency in (\ref{equ:the degradation of optimization problem for throughput maximization}), the instantaneous transmit power for virtual transmitter should satisfy the following conditions.\\
(\,i\,) $p_v(t_0)$ should be either more than $\frac{P_{cons}}{\xi |h_c(t_0)|^2}$ or equal to zero if $t_0\in [-\frac{L_0}{v_0},0]$.\\
(ii) $p_v(t_0)$ should be either more than $\frac{P_{cons}}{\xi |h_c(0)|^2}$ or equal to zero if $t_0\in [0,\frac{L_0}{v_0}]$.}

The proof for Lemma 4.3 is similar to that for Lemma 3.1, which is neglected here. Now, we can reuse the Lagrange multiplier method presented in Section III.B again to derive the expression of $p_v^\ast(t)$. Specifically, the corresponding Lagrange function with respect to $p_v(t)$ can be expressed as
\begin{equation}\label{eqn:the lagrange function in harvest-store-use scheme}
\begin{split}
  \mathfrak{F}_2=& \int_{-\frac{L_0}{v_0}}^{0} \log_2 \Big(1+\Big(\frac{\xi |h_s(\tau)|^2 |h_c(\tau)|^2 p_v(\tau)}{\sigma_0^2} - \frac{|h_s(\tau)|^2 P_{cons}}{\sigma_0^2} \Big)^+\Big)d\tau- \lambda_3 \int_{-\frac{L_0}{v_0}}^{\frac{L_0}{v_0}}p_v(\tau)  d\tau \\
    & \,\,\,\,\,\,\,\,\,\,+\int_{0}^{\frac{L_0}{v_0}}  \log_2 \Big(1+\Big(\frac{\xi |h_s(\tau)|^2 |h_c(0)|^2 p_v(\tau)}{ \sigma_0^2} - \frac{|h_s(\tau)|^2 P_{cons}}{ \sigma_0^2} \Big)^+ \Big) d\tau.
\end{split}
\end{equation}

Since the objective function and constraints are convex, the problem in (\ref{equ:the degradation of optimization problem for throughput maximization}) has a unique solution. By setting the first-order derivative of function $\mathfrak{F}_2$ with respect to $p_v(t)$ be zero, the unique solution can be derived, which is
\begin{equation}\label{eqn:unique optimal solution}
  p_v^\ast(t)=
\begin{cases}
   \Big(\frac{1}{\lambda_2 \ln2}-\frac{d(t)^{(\alpha_s+\alpha_c)}\sigma_0^2}{\xi G_s G_c} +\frac{d(t)^{\alpha_c} P_{cons}}{\xi G_c}\Big)^+,\,t\in [-\frac{L_0}{v_0},0] \\
   \,\,\,\,\Big(\frac{1}{\lambda_2 \ln 2}-\frac{d(t)^{\alpha_s}d_0^{\alpha_c}\sigma_0^2}{\xi G_s G_c} +\frac{d_0^{\alpha_c} P_{cons}}{\xi G_c} \Big)^+\,\,\,\,,\,t\in [\,\,0\,\,,\frac{L_0}{v_0}]
\end{cases}
\end{equation}
where the value of $\lambda_2$ is determined by the constraint (\ref{equ:the degradation of optimization problem for throughput maximization constraint a}).

Substituting (\ref{eqn:unique optimal solution}) into (\ref{equ:the degradation of optimization problem for throughput maximization}), the corresponding maximal cumulative throughput under optimum transmission policy in harvest-store-use scheme can be obtained
\begin{equation}\label{eqn:total throughput with battery}
  R_{\textrm{\tiny OTP}}^{\textrm{\tiny HSU}}=  \int_{-\frac{L_0}{v_0}}^{0} \log_2 \Big[ \max \Big(\tfrac{\xi G_s G_c}{\lambda_2 \ln 2 \cdot d(\tau)^{(\alpha_s+\alpha_c)}\sigma_0^2},1\Big) \Big] d\tau
  +\int_{0}^{\frac{L_0}{v_0}} \log_2 \Big[ \max \Big(\tfrac{\xi G_s G_c}{\lambda_2 \ln2 \cdot d(\tau)^{\alpha_s} d_0^{\alpha_c} \sigma_0^2},1\Big) \Big] d\tau.
\end{equation}

Up to now, $p_v^\ast(t)$ and $R_{\textrm{\tiny OTP}}^{\textrm{\tiny HSU}}$ have been already derived so that the optimum transmission policy $\{p_c^\ast(t)\}$ and $\{p_s^\ast(t)\}$ in (\ref{eqn:optimal power allocation at control center})--(\ref{eqn:optimal power allocation at sensor node}) can be obtained.

\emph{Noting 4.1: Since there is a delta function in the expression of $p_c^\ast(t)$ in (\ref{eqn:optimal power allocation at control center}), it is intuitively difficult to realize in a practical system. From an implementation viewpoint, let $P_m$ be the maximum allowable transmit power at control center, the modified $p_c^\ast(t)$ for a given $P_m$ is}
\begin{equation}\label{eqn:modified optimal power allocation at control center}
  p_c^\ast(t)=
\begin{cases}
   p_v^\ast(t),\,\, t\in [-\frac{L_0}{v_0},-\Delta t_m] \\
   \,\,P_m, \,\,\,\,\, t\in [-\Delta t_m,\,\,\Delta t_m\,] \\
   \,\,\,\,0,\,\,\,\,\,\,\,\,\, t\in [\,\,\,\Delta t_m\,,\,\,\,\frac{L_0}{v_0}\,\,]
\end{cases}
\end{equation}
\emph{where}
\begin{equation}\label{eqn:the parameter of time}
  \Delta t_m=\frac{\int_0^{\frac{L_0}{v_0}} p_v^\ast(\tau)d\tau}{2P_m}.
\end{equation}
\emph{From a perspective of wireless energy transfer, compared with ideal expression in (\ref{eqn:optimal power allocation at control center}), the loss ratio of harvested energy at sensor node resulted from maximal allowable transmit power constraint at control node can be expressed as}
\begin{equation}\label{eqn:normalized performance loss}
  \eta=1-\frac{\int_{-\Delta t_m}^{\Delta t_m} \frac{\xi G_c}{d(\tau)^{\alpha_c}}P_m d\tau}{\int_0^{\frac{L_0}{v_0}} \frac{\xi G_c}{d_0^{\alpha_c}} p_v(\tau)d\tau}.
\end{equation}
\emph{If $P_m$ is big enough, the value of $\Delta t_m$ is relative small. In this case, $d(t) \approx d_0$ when $t\in [-\Delta t_m,\Delta t_m]$. Correspondingly, the energy loss resulted from maximal allowable transmit power $P_m$ is very small, which will also be validated by simulation in Section VI.}

\section{Appropriate Transmission Policy under Fixed Information Rate Constraint}

In Section IV, the optimal transmission policy under the harvest-store-use scheme has been derived in terms of cumulative throughput maximization.
To realize it, an adaptive channel coding is essential for the sensor node, which may lead some difficulties for implementation. In practice, the most important design principle is that the energy harvesting node should be designed as simple as possible, since both the available energy and hardware are limited. Based on above considerations, the transmission policy with fixed information rate constraint (FIRC) at sensors, where the instantaneous data rate is just constant during the transmission period, will be studied in this section. Besides, the small-scale fast fading of wireless channel will also be taken into account as well as the deterministic channel model.

Provided that the transmit power of sensor node at the system time $t$ is $p_s(t)$, the achievable rate profile between the control center and sensor node can be expressed as
\begin{equation}\label{eqn:the instantaneous rate}
  r(t)= \log_2 \big(1+\frac{|h_s(t)|^2 p_s(t)}{\sigma_0^2}\big), \,\,t\in \big[-\frac{L_0}{v_0},\frac{L_0}{v_0}\big].
\end{equation}

In a time-varying channel environment, the truncated transmission profile is beneficial to achieve a better energy efficiency under fixed information rate constraint \cite{Goldsmith_23}. Denote the starting time of data transmission at the sensor node be $t_1$ and the completed time be $t_2$. That is to say, the information is only sent during the active period $[t_1,t_2]$. Otherwise, the sensor keeps silent. In order to guarantee the reliability of information transmission, the practical loaded data rate should be set as the minimum value of $r(t)$ during the whole active period, namely $(t_2-t_1) \cdot \min\{ r(t)|t\in [t_1,t_2] \}$ on the condition that $t_1$ and $t_2$ are given.
Then, the generalized problem for maximizing cumulative throughput with fixed information rate constraint under harvest-store-use scheme can be modeled as follows, which is formulated as
\begin{subequations}\label{equ:optimization problem with constant rate constraint}
\begin{align}
   R_{\textrm{\tiny FIRC}}^{\textrm{\tiny HSU}}&= \max \limits_{p_c(t),p_s(t),t_1,t_2} \big\{|t_2-t_1|\cdot \min \{ r(t)|t\in [t_1,t_2] \} \big\}   \tag{24}\\
   s.t.\,\,\,& \int_{-\frac{L_0}{v_0}}^{t} (p_s(\tau)+P_{cons} I_{\{p_s(\tau)>0\}} ) d\tau \leq \int_{-\frac{L_0}{v_0}}^{t} p_h(\tau)d\tau, \label{equ:optimization problem with constant rate constraint a} \\
   & \int_{-\frac{L_0}{v_0}}^{\frac{L_0}{v_0}} p_c(\tau)d\tau \leq P_0 \cdot \frac{2L_0}{v_0}, \label{equ:optimization problem with constant rate constraint b} \\
   & p_c(t), p_s(t) \geq 0 ,t\in [-\tfrac{L_0}{v_0},\tfrac{L_0}{v_0}].  \label{equ:optimization problem with constant rate constraint d}
\end{align}
\end{subequations}
where the inequality in (\ref{equ:optimization problem with constant rate constraint a}) is the causality constraint for energy consumption at sensor node, the equality in (\ref{equ:optimization problem with constant rate constraint b}) denotes the power constraint at control node and the equality in (\ref{equ:optimization problem with constant rate constraint d}) reflects the nonnegativity of transmit power.

Let $r_0=(t_2-t_1)\cdot \min\{ r(t)|t\in [t_1,t_2]\}$, the most efficient expression for $r(t)$ in terms of energy efficiency maximization should guarantee no energy wasting, and it can be expressed as
\begin{equation}\label{eqn:expression for instantaneous information rate}
  r(t)=
\begin{cases}
   \frac{r_0}{t_2-t_1},\,\, t\in [t_1,t_2] \\
   \,\,\,\,\,0\,\,\,\,,\,\, t \in [-\frac{L_0}{v_0},t_1]\cup[t_2,\frac{L_0}{v_0}]
\end{cases}
\end{equation}

Combining (\ref{eqn:the instantaneous rate}) and (\ref{eqn:expression for instantaneous information rate}), the corresponding $p_s(t)$ that can satisfy the constraint in (\ref{eqn:expression for instantaneous information rate}) is
\begin{equation}\label{eqn:the expression for transmit power at sensors}
  p_s(t)=
\begin{cases}
   \,\frac{1}{\xi |h_s(t)|^2}(2^{\frac{r_0}{t_2-t_1}}-1),\,\,\, t\in [t_1,t_2] \\
   \,\,\,\,\,\,\,\,\,0\,\,\,\,\,\,\,,\,\, t \in [-\frac{L_0}{v_0},t_1]\cup[t_2,\frac{L_0}{v_0}]
\end{cases}
\end{equation}

\emph{Lemma 5.1: Under optimal transmission policy, the cumulative throughput $r_0^\ast$ is a monotone-increasing function with respect to the average transmit power at control center $P_0$.}
\begin{IEEEproof}
It is easy to see that a larger $P_0$ amounts to looser constraint in (\ref{equ:optimization problem with constant rate constraint b}), so that a larger optimal value can be achieved \cite{Boyd_28}. Thus, the Lemma 5.1 has been proved.
\end{IEEEproof}

Based on Lemma 5.1, maximizing $r_0$ with fixed $P_0$ constraint is equivalent to minimizing $P_0$ with fixed $r_0$ constraint. As a result, the problem in (\ref{equ:optimization problem with constant rate constraint}) can be reformulated as
\begin{subequations}\label{equ:rewritten optimization problem with constant rate constraint}
\begin{align}
   P_0= & \min \limits_{p_c(t),t_1,t_2} \mathbb{E} \Big\{\frac{v_0}{2L_0} \int_{-\frac{L_0}{v_0}}^{\frac{L_0}{v_0}} p_c(\tau) d\tau\,\,\Big\}   \tag{27}\\
   s.t.\,\,\,& \int_{-\frac{L_0}{v_0}}^{t} (p_s(\tau)+P_{cons} I_{\{p_s(\tau)>0\}} ) d\tau \leq \int_{-\frac{L_0}{v_0}}^{t} p_h(\tau)d\tau, \label{equ:rewritten optimization problem with constant rate constraint a} \\
   &   p_s(t)=
\begin{cases}
   \,\frac{1}{\xi |h_s(t)|^2}(2^{\frac{r_0}{t_2-t_1}}-1),\,\,\, t\in [t_1,t_2] \\
   \,\,\,\,\,\,\,\,\,0\,\,\,\,\,\,\,,\,\, t \in [-\frac{L_0}{v_0},t_1]\cup[t_2,\frac{L_0}{v_0}]
\end{cases} \label{equ:rewritten optimization problem with constant rate constraint b} \\
   & p_c(t) \geq 0 ,t\in [-\tfrac{L_0}{v_0},\tfrac{L_0}{v_0}].  \label{equ:rewritten optimization problem with constant rate constraint d}
\end{align}
\end{subequations}
where $\mathbb{E}\{\cdot\}$ denotes the statistical mean operation for the channel state variables.

\subsection{Deterministic Channel Model}

Let us consider above problem under the deterministic channel model, where $h_c(t)=\sqrt{\frac{G_c}{d(t)^{\alpha_c}}}$ and $h_s(t)=\sqrt{\frac{G_s}{d(t)^{\alpha_s}}}$. Firstly, we need to relax the causality constraint in (\ref{equ:optimization problem with constant rate constraint a}), then attempting to obtain an explicit expression of optimum transmission policy under fixed information rate constraint. Similar to the discussion in Section IV, it is assumed that there is a virtual energy transmitter transferring energy with power $p_v(t)$ from control center to sensor node, the average value of which is equal to that of $p_c(t)$ during a whole transmission period. Namely
\begin{equation}\label{eqn:constraint for virtual transmit power}
  \int_{-\frac{L_0}{v_0}}^{\frac{L_0}{v_0}} p_v(\tau)d\tau = \int_{-\frac{L_0}{v_0}}^{\frac{L_0}{v_0}} p_c(\tau)d\tau.
\end{equation}

These results in Lemma 4.1 and Lemma 4.2 are still tenable in this case. So we can establish the relationship between $p_v(t)$ and the required data rate $r_0$ by combining (\ref{eqn:optimal power allocation at sensor node}) and (\ref{equ:rewritten optimization problem with constant rate constraint b})
\begin{equation}\label{eqn:the expression for virtual transmit power}
  P_v(t)=
\begin{cases}
   \,\frac{d(t)^{\alpha_s+\alpha_c}}{\xi G_s G_c}(2^{\frac{r_0}{t_2-t_1}}-1)+\frac{d(t)^{\alpha_c}}{\xi G_c}P_{cons},\,\,\, t\in [t_1,0] \\
   \,\,\,\frac{d(t)^{\alpha_s}d_0^{\alpha_c}}{\xi G_s G_c}(2^{\frac{r_0}{t_2-t_1}}-1)+\frac{d_0^{\alpha_c}}{\xi G_c}P_{cons}\,\,\,,\,\,\, t\in [0,t_2] \\
   \,\,\,\,\,\,\,\,\,0\,\,\,\,\,\,\,\,,\,\, t \in [-\frac{L_0}{v_0},t_1]\cup[t_2,\frac{L_0}{v_0}]
\end{cases}
\end{equation}

Substituting (\ref{eqn:constraint for virtual transmit power})--(\ref{eqn:the expression for virtual transmit power}) into (\ref{equ:rewritten optimization problem with constant rate constraint}), it yields
\begin{equation}\label{eqn:constraint for virtual transmit power 2}
\begin{split}
  P_0= & \min \limits_{t_1,t_2} \mathfrak{L}(r_0,t_1,t_2)=
  \min \limits_{t_1,t_2} \Big\{ \frac{v_0(2^{\frac{r_0}{t_2-t_1}}-1)}{2 L_0 \xi G_s G_c}
  \big[\int_{t_1}^{0}d(\tau)^{\alpha_s+\alpha_c} d\tau+d_0^{\alpha_c} \int_{0}^{t_2}d(\tau)^{\alpha_s} d\tau \big] \\
  &+\frac{P_{cons}}{\xi G_c}\int_{t_1}^{0}d(\tau)^{\alpha_c}d\tau+\frac{P_{cons}}{\xi G_c} d_0^{\alpha_c} t_2 \Big \}.
\end{split}
\end{equation}

It is easy to see that the only residual task that needs to do for obtaining optimum transmission policy is to calculate the minimum value of $P_0$, by optimizing it with respect to $t_1$ and $t_2$, when $r_0$ is given.
After obtaining the optimal time-domain truncated parameters $t_1^\ast$ and $t_2^\ast$, the optimal profile $\{p_v^\ast(t)\}$ can be obtained by substituting these parameters into (\ref{eqn:the expression for virtual transmit power}). Then, the optimum transmission policy under fixed information rate constraint, namely $\{p_c^\ast(t)\}$ and $\{p_s^\ast(t)\}$, can be derived by substituting $\{p_v^\ast(t)\}$ into (\ref{eqn:optimal power allocation at control center}-\ref{eqn:optimal power allocation at sensor node}).

Now let us study the relationship between the values of $t_1$ and $t_2$ under optimum transmission policy, by which one of $t_1$ and $t_2$ can be eliminated, resulting in low complexity to solve the optimization problem in (\ref{eqn:constraint for virtual transmit power 2}). Recall that $t=0$ is the most efficient opportunity for information and energy transfer, the effective time window $[t_1,t_2]$ must contain the point $t=0$. That is to say, $-\frac{L_0}{v_0} \leq t_1 \leq 0 \leq t_2 \leq \frac{L_0}{v_0}$.

\emph{Lemma 5.2: Let $t_1^\ast$ and $t_2^\ast$ be the optimal boundary points with respect to fixed information rate constraint under optimal transmission policy. For maximizing energy efficiency, the equivalent propagation loss at system time $t_1^\ast$ should be equal to that at system time $t_2^\ast$. Namely,}
\begin{equation}\label{eqn:relationship between equivalent propagation loss two times}
  \frac{\xi G_s G_c}{d(t_1^\ast)^{\alpha_s+\alpha_c}}=\frac{\xi G_s G_c}{d(t_2^\ast)^{\alpha_s}d_0^{\alpha_c}},-\frac{L_0}{v_0} \leq t_1 \leq 0 \leq t_2 \leq \frac{L_0}{v_0}.
\end{equation}
\begin{IEEEproof}
 See the Appendix E.
\end{IEEEproof}

By solving the Equation in (\ref{eqn:relationship between equivalent propagation loss two times}), $t_1^\ast$ can be expressed as a function of $t_2^\ast$, which is
\begin{equation}\label{eqn:t-1 is a function of t-2}
  t_1^\ast=\mathfrak{T}(t_2^\ast)=\frac{  \Big\{\big[d_0^{\alpha_c} \cdot d(t_2^\ast)^{\alpha_s} \big]^{\frac{2}{\alpha_s+\alpha_c}}-d_0^2 \Big\}^{\frac{1}{2}} }{v_0}
\end{equation}

Substituting the result in (\ref{eqn:t-1 is a function of t-2}) into (\ref{eqn:constraint for virtual transmit power 2}), the variable $t_1$ can be eliminated. After that
\begin{equation}\label{eqn:constraint for virtual transmit power 3}
  P_0= \min \limits_{t_2} \,\{\mathfrak{L}(r_0,\mathfrak{T}(t_2),t_2)\}.
\end{equation}

Thus, when the value of $r_0$ is given, the operation for minimizing the value of $P_0$ in (\ref{eqn:constraint for virtual transmit power 2}) is only needed to optimize over all possible $t_2$ (or $t_1$), in which just one dimension searching is needed. Based on it, the corresponding optimal truncated parameters $(t_1^\ast, t_2^\ast)$ can be obtained. Up to now, the problem in (\ref{equ:optimization problem with constant rate constraint}) has been completely solved in a deterministic channel model. Specifically, $\{p_v^\ast(t)\}$ can be derived by substituting $(t_1^\ast, t_2^\ast)$ and $r_0$ into (\ref{eqn:the expression for virtual transmit power}). And the optimum transmission policy in this case can be derived by substituting $\{p_v^\ast(t)\}$ into (\ref{eqn:optimal power allocation at control center}--\ref{eqn:optimal power allocation at sensor node}).

\subsection{Random Fading Channel Model}

In this subsection, we further study the optimization problem in (\ref{equ:rewritten optimization problem with constant rate constraint}) under a random fading channel model, in which both small-scale fading and large-scale fading are considered together. We assume that $h_c(t)=h_s(t)=h(t)$ in a symmetric system architecture, in which
\begin{equation}\label{equ:random fading channel model}
  h(t)=\sqrt{\frac{G}{d(t)^{\alpha}}} \beta(t),\,\,t\in \big[-\frac{L_0}{v_0},\frac{L_0}{v_0}\big],
\end{equation}
where $\beta(t)$ represents the small-scale fading coefficient, the amplitude of which satisfies the Nakagami $m$ distribution with unit variance. $G$ and $\alpha$ denote the constant power gain and the path loss exponent, respectively.

Let us consider the optimal opportunity for wireless energy transfer in this condition.
It is assumed that the channel side information is causally known by the system via particular estimation methods (see e.g., \cite{Zhang_29,Zeng_30}).
Due to the randomness of $\beta(t)$, the statistical mean operation in (\ref{equ:rewritten optimization problem with constant rate constraint}) can not be ignored at all. So, we can not get certain deterministic result for the optimal opportunity of wireless energy transfer presented in Lemma 4.1 and 4.2.
However, these results are still heuristic for the problem solving in this new case. Similar to the analysis in Section IV.B, let $p_p(t_0) \Delta t$ be the energy consumption by sensor node at system time $t_0$. Then, we discuss when is the best opportunity to transfer these energy $p_p(t_0) \Delta t$ from control center to sensor node. Some conclusions are presented as follows.

\emph{Lemma 5.3: When $t_0\in[-\frac{L_0}{v_0},0]$, in terms of the first-order statistics, $t=t_0$ is the best opportunity to transfer the energy $p_p(t_0)\Delta t$ that is consumed at system time $t_0$ from control center to sensor node. Otherwise, $t=0$ is the best opportunity to transfer the energy $p_p(t_0)\Delta t$ from control center to sensor node.}
\begin{IEEEproof}
 See the Appendix F.
\end{IEEEproof}

\begin{algorithm}[htb]
\caption{A heuristic algorithm for the system under the random fading scenario}\label{alg:Framwork}
\begin{algorithmic}[1] 
\STATE For the value of $r_0$, calculating $p_c(t)$ and $p_p(t)$ by the methods developed in Section V.A;
\STATE Based on observed value of $\beta(t)$, calculating the real-time energy consumption requirement $p_p'(t)$ at sensor node to support $r_0$, as shown in (\ref{equ:required energy consumption});
\IF{$p_p'(t) \leq p_p(t)$}
\STATE It means that the proactively scheduled energy is more than the required, sensor node can save the left energy $p_p(t)-p_p'(t)$ into a subset of the energy buffer, denoted as $Q$;
\ELSE
\IF{There is sufficient energy in $Q$ to support $p_p'(t)-p_p(t)$}
\STATE The sensor uses the energy in $Q$ to cover the current balance $p_p'(t)-p_p(t)$;
\ELSE
\STATE The control center increases current transmit power to complement $p_p'(t)-p_p(t)$.
\ENDIF
\ENDIF
\end{algorithmic}
\end{algorithm}

According to Lemma 5.3, the control center can get a power-bearing signal profile $\{p_c(t)|t\in [-\frac{L_0}{v_0},\frac{L_0}{v_0}]\}$ based on the results that have been derived in the deterministic channel model. Correspondingly, the sensor can get an available energy file for feedback signal transmission, denoted as $\{p_p(t)|t\in [-\frac{L_0}{v_0},\frac{L_0}{v_0}]\}$. On the other hand, based on the causally known channel information, the real energy consumption requirement for supporting the stable data flow is
\begin{equation}\label{equ:required energy consumption}
  p_p'(t)=
\begin{cases}
   \,\frac{d(t)^{\alpha}}{\xi G |\beta(t)|^2}(2^{\frac{r_0}{t_2-t_1}}-1)+P_{cons},\,\,\, t\in [t_1,t_2] \\
   \,\,\,\,\,\,\,\,\,0\,\,\,\,\,\,\,,\,\, t \in [-\frac{L_0}{v_0},t_1]\cup[t_2,\frac{L_0}{v_0}]
\end{cases}
\end{equation}

Due to the randomness of variable $\beta(t)$, it is intuitive that sometimes $p_p'(t)>p_p(t)$ and sometimes $p_p'(t)\leq p_p(t)$. In order to tackle this mismatching, a specifical near-optimal algorithm is proposed to instruct the system how to deliver the energy in an efficient way, which is presented in Algorithm \ref{alg:Framwork}. Since there is no an explicit expression for the result, the system performance will be discussed via simulation in the next section.

\section{Simulation Results}

This section will validate the theoretical results developed in this paper by simulation. For the unit system shown in Fig. 1(b), the common part of system parameters involved in the simulation are given in TABLE \ref{table one}.
\begin{table}
\centering
\caption{The system parameter setting in the simulation.}\label{table one}
\begin{tabular}{|c|c|}
\hline
SYSTEM PARAMETERS  &  VALUE \\
\hline
Constant power gain of two links $G_c$ and $G_s$  & 10 dB \\
\hline
Path loss exponent $\alpha_c$ and $\alpha_s$ & 3  \\
\hline
Baseband circuit power consumption $P_{cons}$ & 5 mW \\
\hline
The moving velocity of control center $v_0$ & 20 m/s \\
\hline
The coefficient of energy harvesting efficiency $\xi$ & 50\%\\
\hline
The radius of effective coverage of each sensor $|OA|=|OC|=d_m$ in Fig. \ref{fig:system_structure for unit} & $\sqrt{100^2+10^2}$ m \\
\hline
The distance between sensor and moving track $|OB|=d_0$ in Fig. \ref{fig:system_structure for unit} & 10 m \\
\hline
The moving range $|AB|=|BC|=L_0$ in Fig. \ref{fig:system_structure for unit} & 100 m \\
\hline
\end{tabular}
\end{table}
Fig. 3(a) depicts the cumulative throughput (in bit/Hz) that can be achieved between the mobile control center and sensor node during a whole period as a function of average signal to noise ratio (SNR), namely $\frac{P_0}{\sigma_0^2}$. Three specifical transmission strategies are considered in Fig. 3(a), which contains CTP/OTP in harvest-and-use scheme and OTP in harvest-store-use scheme. The performance of CTP in harvest-and-use scheme, which is widely used in quasi-static and short-distance systems, is served as a baseline.
By allocating transmit power adaptively at mobile control center based on optimization theory, OTP in harvest-and-use scheme can obtain a better performance, especially in the lower SNR range. When SNR goes to infinite, CTP and OTP in harvest-and-use scheme tend to the same.
On the other hand, if the energy transfer and information transmission are jointly optimized, significant improvement can be observed.
For instance, in order to obtain $60$ bit/Hz total throughput within $20$s, the improvement is about $10$ dB in terms of energy efficiency, which means that more than $90\%$ energy at control center can be saved compared with traditional two other strategies. Moreover, the modified OTP case for energy profile at control center in (\ref{eqn:modified optimal power allocation at control center}) is also considered, where the maximum allowable power is $\frac{P_m}{P_0}=20$ dB. It can be observed from Fig. 3(a) that the degradation of throughput is not apparent compared with ideal OTP case in (\ref{eqn:optimal power allocation at control center}).

The relationship between system performance and propagation attenuation exponent is also investigated in this section.
It is assumed that the two links in Fig. 1(b) satisfy $\alpha_c=\alpha_s=\alpha_0$. Fig. 3(b) plots the overall energy consumption normalized to that under CTP in harvest-and-use scheme as a function of propagation loss exponent $\alpha_0$, when the throughput requirement is $60$ bit/Hz during a whole transmission period. Similar to that in Fig. 3(a), three specifical transmission policies are taken into account in Fig. 3(b), which are CTP/OTP in harvest-and-use scheme and OTP in harvest-store-use scheme.
It can be observed that the huge improvement can be achieved with the help of optimal transmission policy. In particular, when the value of $\alpha_0$ is big, the propagation attenuation is time-varying rapidly due to the mobility of control center, which is more beneficial for employing opportunistic energy and information transfer strategy.

%

\begin{figure}[!t]
\centering
\subfloat[]{
\label{fig:simulation result 1}
\begin{minipage}[t]{0.5\textwidth}
\centering
\includegraphics[width=3.0 in]{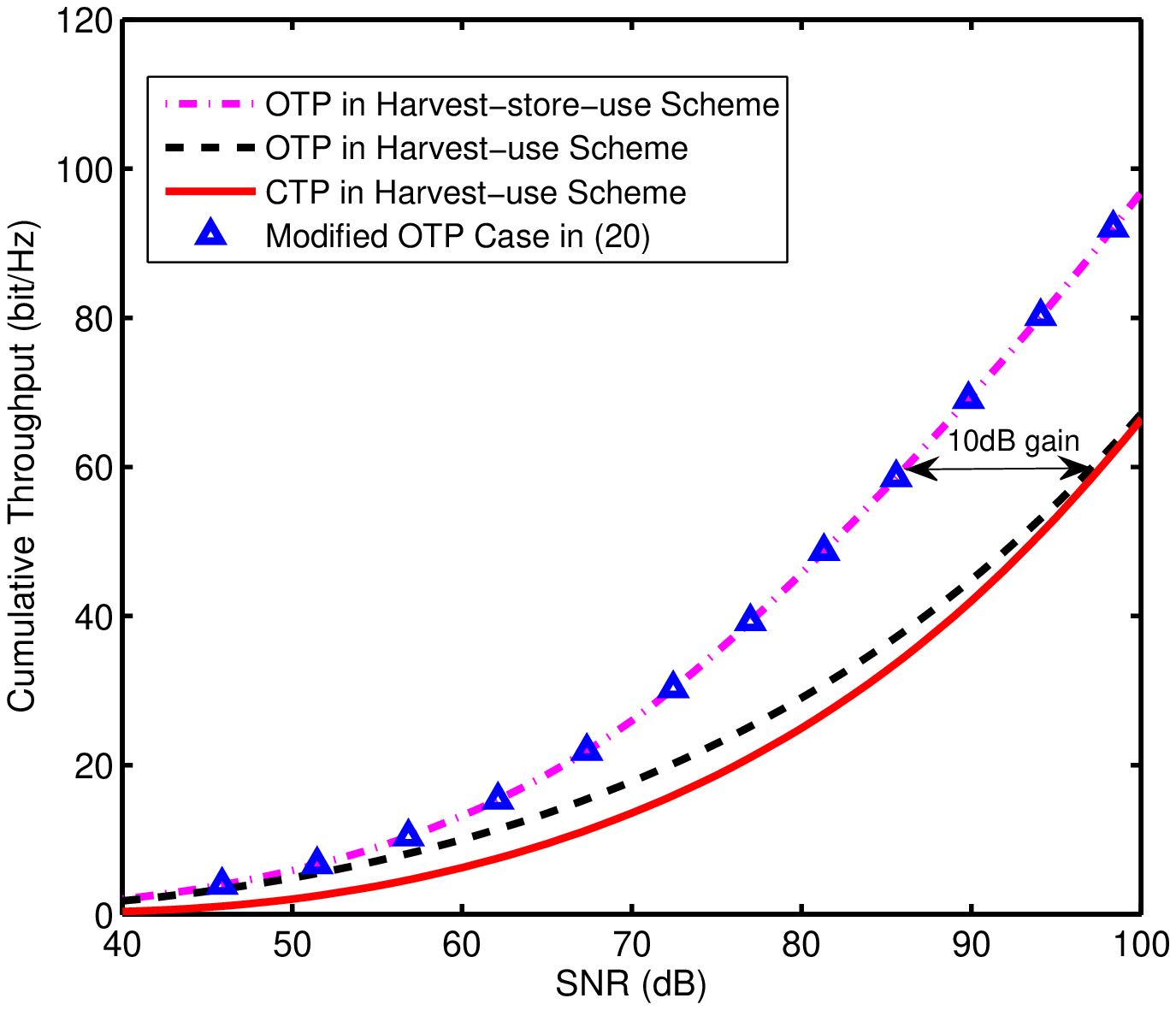}
\end{minipage}
}
\subfloat[]{
\label{fig:simulation result 2}
\begin{minipage}[t]{0.5\textwidth}
\centering
\includegraphics[width=3.0 in]{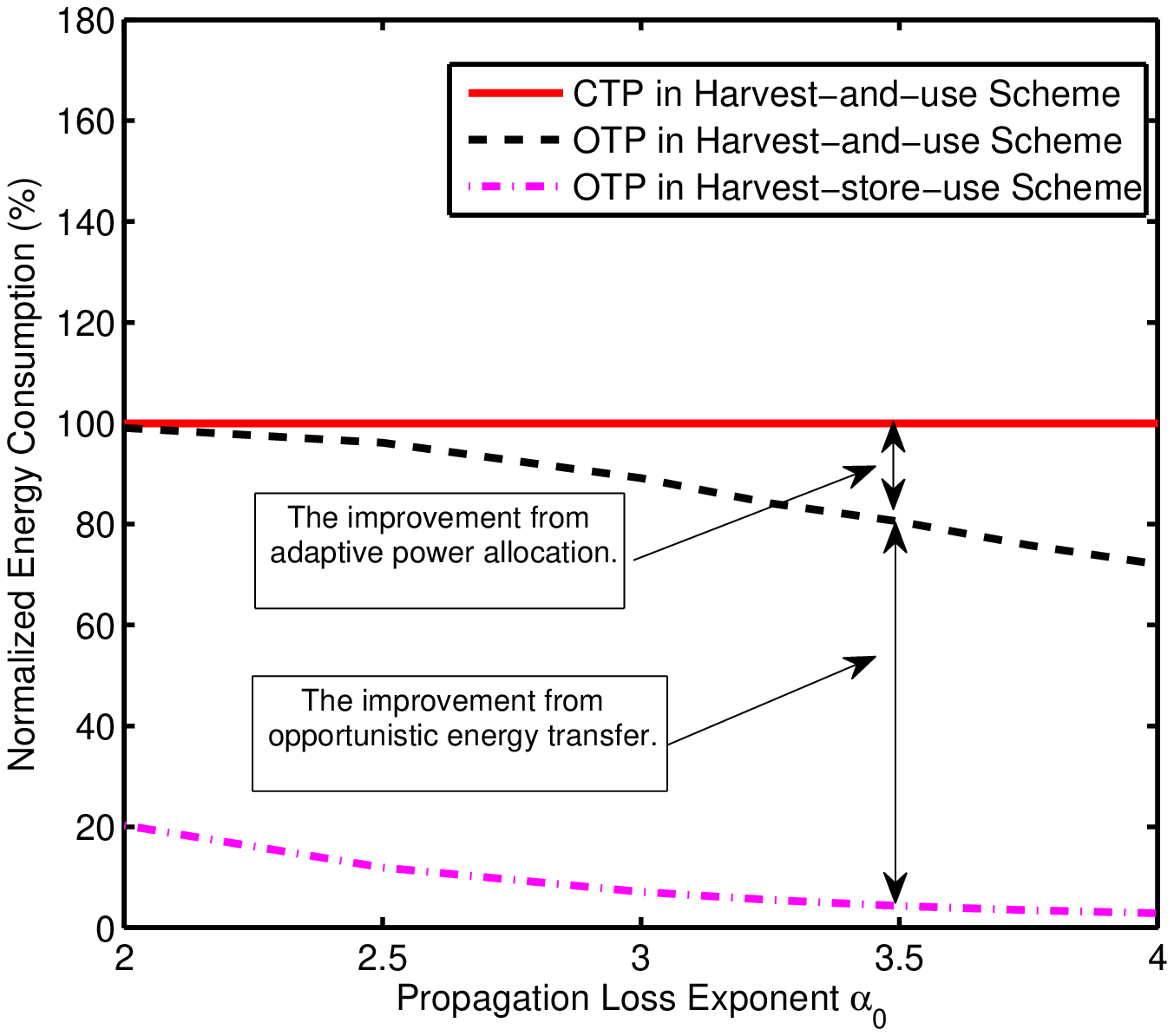}
\end{minipage}
}
\caption{The comparison between three transmission policies: (a) the relation of cumulative throughput and SNR, (b) the relation of energy consumption and path loss exponent.}
\end{figure}

\begin{figure}[!t]
\centering
\includegraphics[width=3.0 in]{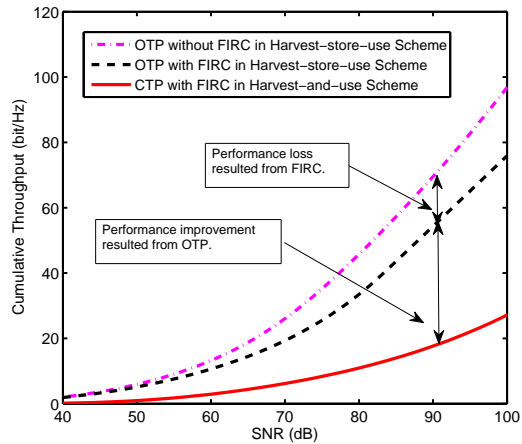}
\caption{The cumulative throughput (in bit/Hz) during a whole period in three different transmission policies with/without FIRC as a function of SNR.}
\label{fig:simulation result 3}
\end{figure}

Fig. \ref{fig:simulation result 3} plots the cumulative throughput under optimal transmission policy with fixed information rate constraint (OTP-FIRC) in harvest-store-use scheme as a function of SNR. As a comparison, the performance of OTP without fixed information rate constraint in (\ref{eqn:optimal power allocation at control center}-\ref{eqn:optimal power allocation at sensor node}) are also given, which can be regarded as an upper bound. Besides, the performance of CTP under fixed information rate constraint (CTP-FIRC) in harvest-and-use scheme is also provided to serve as a baseline. It can be seen that the improvement of optimal policy is relative large compared with CTP-FIRC in harvest-and-use scheme, while the degradation caused by fixed information rate constraint is relatively small. For instance, when the SNR is $90$dB, the improvement resulted from optimal strategy is more than $200\%$ in terms of cumulative throughput while the performance loss resulted from fixed rate constraint is less than $20\%$. Through there is some performance loss in OTP-FIRC scheme, the system can adopt an fixed channel coding mapping strategy to achieve a consistent reliability requirement, which is meaningful in practical system design.

Lastly, let us evaluate the system performance of our strategies in a random fading channel environment, the randomness of which is characterized by Nakagami $m$ distribution.
Fig. 5(a) depicts the average power requirement from a view of control center as a function of cumulative throughput with FIRC, where $\sigma_0^2=1$ and $m=3,6,50,+\infty$, respectively. In particular, $m=+\infty$ corresponds to the case without random fading, namely deterministic channel model discussed in Section V.A. As observed from Fig. 5(a), it is apparent that the results derived in deterministic channel model can be regarded as an upper bound for the system under random fading channel. And the uncertainty of fading can lead to some performance loss to the system.
The smaller the value of $m$ is, the more serious the uncertainty is and the worse the transmission performance is. For instance, when $m=6$ and cumulative throughput is $60$ bit/Hz (during one circle), the performance loss in terms of power consumption is about $15$ dB compared with the upper bound.

The performance of the proposed strategy is further evaluated in Fig. 5(b) by comparing with the traditional strategy, i.e., the CTP with FIRC in the harvest-and-use scheme that has been discussed in Fig. 3--4. Three different fading channel scenarios are considered, which correspond to $m=3,6,20$. It can be seen that the uncertainty of the fading can decrease the benefits brought from optimal transmission strategy. Namely, for the data requirement $R_0=70$ bit/Hz (during one circle), the performance gain is nearly $8$ dB when $m=20$, while the gain is just about $1.5$ dB when $m=3$. However, the performance of it is still better than that of the traditional strategy with the same channel environment, which validates the results developed in this paper.

%

\begin{figure}[!t]
\centering
\subfloat[]{
\label{fig:simulation result 4}
\begin{minipage}[t]{0.5\textwidth}
\centering
\includegraphics[width=3.0 in]{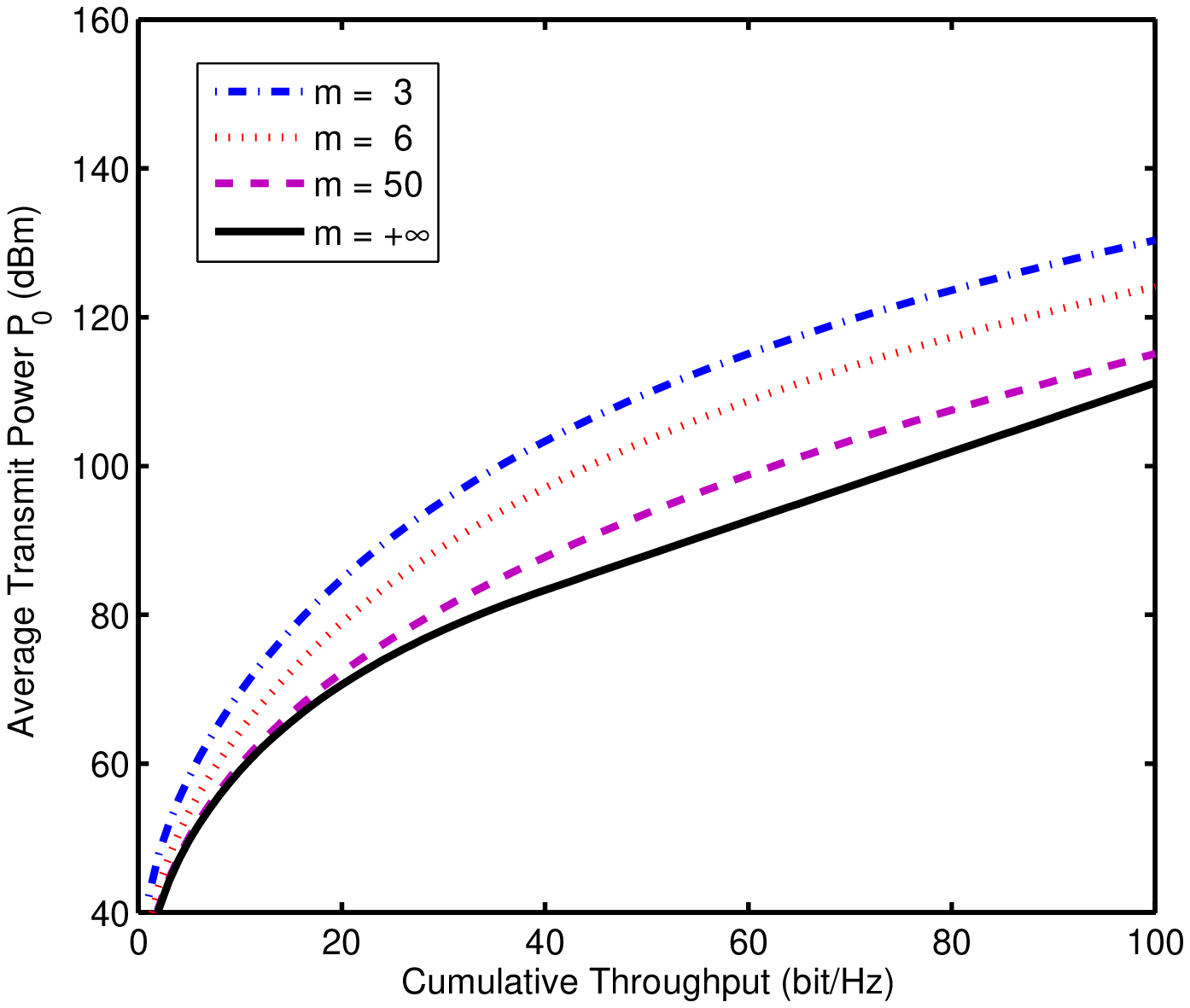}
\end{minipage}
}
\subfloat[]{
\label{fig:simulation result 5}
\begin{minipage}[t]{0.5\textwidth}
\centering
\includegraphics[width=3.0 in]{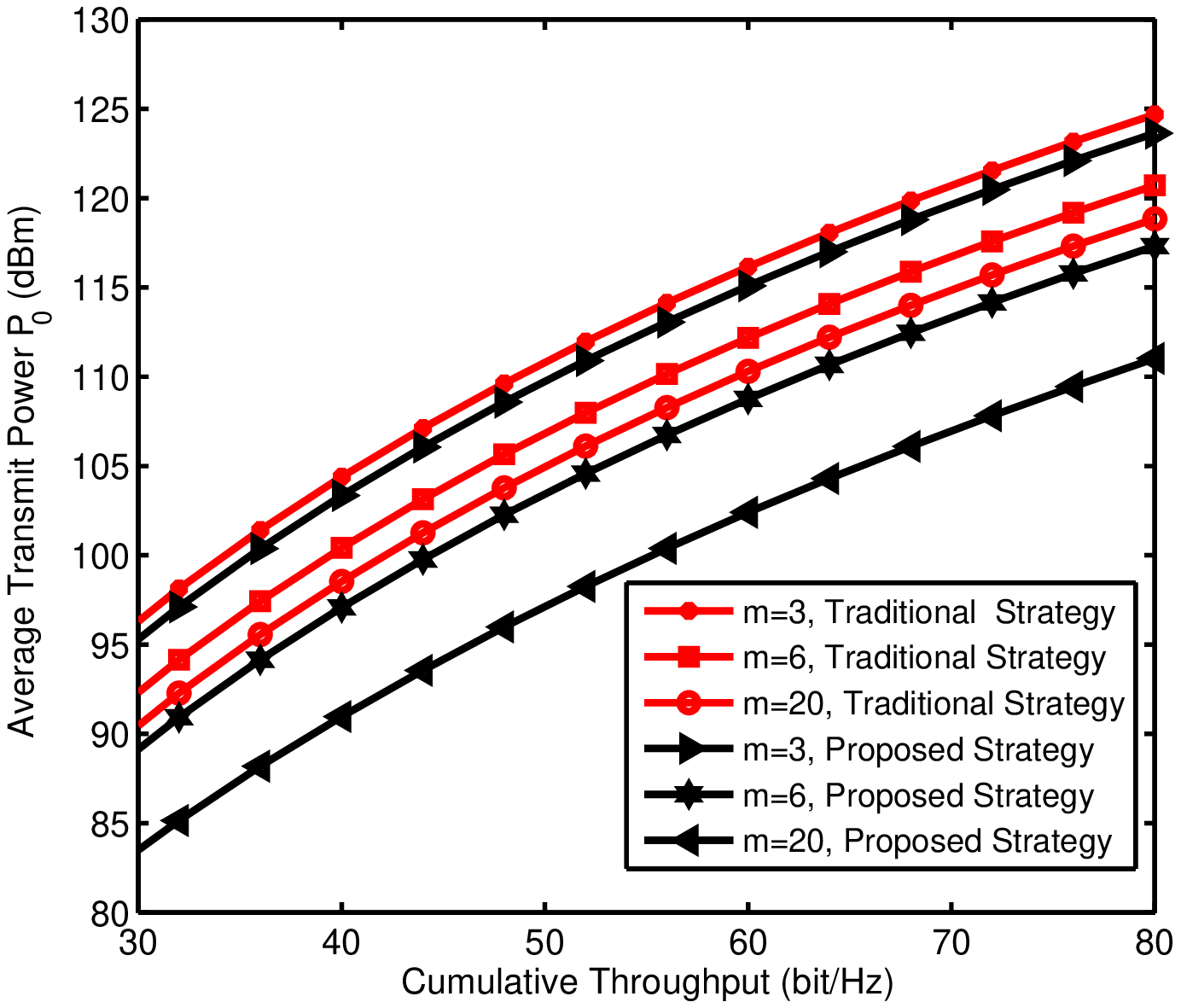}
\end{minipage}
}
\caption{The performance of the new proposed strategy in a Nakagami fading channel: (a) the effect of the uncertainty of fading, (b) The performance comparison with the traditional scheme.}
\end{figure}

\section{Conclusion}

For wireless sensor networks powered by RF-based energy harvesting in a mobile scenario, energy efficiency is the biggest bottleneck of large-scale applications, which motivates us to explore the optimum transmission policy to achieve the most efficient performance of system.
As a baseline, we firstly investigated the throughput performance of constant transmission policy in harvest-and-use scheme, where the harvested energy is utilized immediately after it has been harvested at sensors. Then, the corresponding optimum adaptive transmission policy in harvest-and-use scheme was also studied. After that, we concentrated on the transmission policies in harvest-store-use scheme, where the sensor can store some energy temporarily to achieve a better performance by opportunistic wireless energy transfer. By exploring the best opportunity of wireless energy transfer, the optimum transmission policy in this condition was given under which significant improvement in terms of energy efficiency can be obtained compared with other conventional policies. Namely, a considerable portion of energy at control center can be saved by employing appropriate transmission policy. At last, the transmission policy under fixed information rate constraint was also discussed for its low implementation complexity. In addition, a specific near optimal algorithm was given for the system under random fading environment.


\section*{Appendix A: Proof of Lemma 3.1}

Let us prove it by the contradiction. Assuming the optimal policy $p_c^\ast(t)$ at system time $t_0$ satisfies $0<p_c^\ast(t_0) \leq \frac{P_{cons}}{\xi |h_c(t)|^2}$, then the corresponding harvested energy at $t_0$ is $p_h(t_0) \leq P_{cons}$. Since the available energy $p_h(t_0)$ at current moment $t=t_0$ is less than threshold value for activating the sensor node to feedback data, it means the energy harvested at this moment will be wasted due to the absence of energy storage.
However, if we reload this part of energy at particular time segment $t_0'$ that the sensor node has been activated, it is intuitive that more throughput can be obtained due to the increment of available transmit energy at time segment $t_0'$, which is contradicted with the assumption that $p_c^\ast(t)$ is optimal. As a result, $p_c(t)$ under optimal policy should be either more than $\frac{P_{cons}}{\xi |h_c(t)|^2}$ or equal to zero.

\section*{Appendix B: Proof of Proposition 3.1}

With the help of Lemma 3.1, it can be proved that both the objective function and constraints in (\ref{equ:optimization problem for throughput maximization without battery}) are convex, so that the problem has a unique solution. And the constraint in (\ref{equ:optimization problem for throughput maximization without battery b}) is satisfied with equality at optimal solution, which means no energy wasting at sensor node. Let us define a Lagrange multiplier function as follows, in which $\lambda_1$ is a under-determined constant parameter,
\begin{equation}\label{eqn:the lagrange function in harvest-use scheme}
  \mathfrak{F}_1=
  \int_{-\frac{L_0}{v_0}}^{\frac{L_0}{v_0}} \log_2 \Big(1+\Big(\frac{\xi G_s G_c P_c(\tau)}{d(\tau)^{\alpha_s+\alpha_c} \sigma_0^2} - \frac{G_s P_{cons}}{d(\tau)^{\alpha_s} \sigma_0^2} \Big)^+\Big) d\tau
   -\lambda_1 \int_{-\frac{L_0}{v_0}}^{\frac{L_0}{v_0}}P_c(\tau)  d\tau.
\end{equation}

Combining the results in Lemma 3.1, and setting the first-order derivative of function $\mathfrak{F}_1$ to zero with respect to the variable $p_c(t)$, we can get the result in (\ref{eqn:optimal power allocation without battery}).

\section*{Appendix C: Proof of Lemma 4.1}

For maximizing energy efficiency, the basic idea is to avoid large propagation attenuation as much as possible without affecting normal energy consumption at sensor node. According to (\ref{eqn:transmission distance}) and (\ref{eqn:harvested energy at sensor node}), under the deterministic channel model, the propagation attenuation for wireless energy transfer from control center to sensor node at the system time $t$ can be expressed as
\begin{equation}\label{eqn:propagation attenuation for energy transfer}
  L(t)= \frac{\xi G_c}{(d_0^2+(v_0t)^2)^{\frac{\alpha_c}{2}}}, \, t \in \Big[-\frac{L_0}{v_0}, \frac{L_0}{v_0}\Big].
\end{equation}

Due to the causality constraint at sensor node, the energy consumed at system time $t_0$ has to be transferred from control center to sensor node during the range $[-\frac{L_0}{v_0},t_0]$. For $t_0\in[-\frac{L_0}{v_0},0]$, the time $t=t_0$ is the best opportunity among the effective range $t\in [-\frac{L_0}{v_0},t_0]$ for energy transfer based on (\ref{eqn:propagation attenuation for energy transfer}) in terms of maximizing energy efficiency, since the energy propagation attenuation $L(t)$ at $t_0$ is the smallest among the whole available range $[-\frac{L_0}{v_0},t_0]$. 

\section*{Appendix D: Proof of Lemma 4.2}

Similar to the proof of Lemma 4.1, when $t_0\in[0,\frac{L_0}{v_0}]$, it can be obtained based on (\ref{eqn:propagation attenuation for energy transfer}) that the time $t=0$ is the best opportunity for wireless energy transfer from central to sensor node to achieve the largest energy efficiency, since the propagation attenuation $L(t=0)$ is the smallest among the available range $[-\frac{L_0}{v_0},t_0]$. Thus, the energy consumed by sensor node at time $t_0$ should be transferred at time $t=0$.

\section*{Appendix E: Proof of Lemma 5.2}

For the whole transmission period $t \in [-\frac{L_0}{v_0},\frac{L_0}{v_0}]$, with the help of virtual transmitter based on opportunistic wireless energy transfer, the equivalent propagation attenuation according to Lemma 4.1 and Lemma 4.2 can be expressed as
\begin{equation}\label{eqn:equivalent propagation attenuation in CRC scheme}
  L_e'(t)=
\begin{cases}
   \frac{\xi G_s G_c}{d(t)^{\alpha_s+\alpha_c}},\,\, t\in [-\frac{L_0}{v_0},0] \\
   \frac{\xi G_s G_c}{d(t)^{\alpha_s}d_0^{\alpha_c}},\,\, t \in [\,\,0\,\,,\frac{L_0}{v_0}].
\end{cases}
\end{equation}

It is observed that $L_e'(t)$ is monotone-decreasing function with respect to the system time $t$ when  $t\in [-\frac{L_0}{v_0},0]$, while $L_e'(t)$ is monotone-increasing function when  $t\in [0,\frac{L_0}{v_0}]$. We will prove this conclusion by the contradiction. Without loss of generality, it is assumed that $L_e'(t_1^\ast)<L_e'(t_2^\ast)$ under optimal transmission policy, similar results can be extended for the case $L_e'(t_1^\ast)>L_e'(t_2^\ast)$. Let $\Delta t$ be an arbitrary small time segment. According to (\ref{eqn:equivalent propagation attenuation in CRC scheme}), the time fragment $[t_1^\ast-\Delta t,t_1^\ast]$ is more efficient in terms of energy efficiency than the time fragment $[t_2^\ast-\Delta t,t_2^\ast]$. Thus, $t_1^\ast-\Delta t$ and $t_2^\ast-\Delta t$ are more appropriate for improving the system performance, which is contradicted with the assumption that $t_1^\ast$ and $t_2^\ast$ is optimal. As a result, the conclusion in (\ref{eqn:relationship between equivalent propagation loss two times}) is obtained.

\section*{Appendix F: Proof of Lemma 5.3}

During the period of $[-\frac{L_0}{v_0},0]$, the channel propagation attenuation consists of two parts: large-scale fading $\frac{G}{d(t)^{\alpha}}$ and small-scale fading $\beta(t)$. Since $\beta(t)$ is an ergodic stochastic process and is only causally known by the control center, we can not make sure deterministically which point among $[-\frac{L_0}{v_0},0]$ is the best opportunity for transferring. However, It can be assured that $t=t_0$ is the optimal point in terms of first-order statistics due to the ergodic property of $\beta(t)$. Thus, $t=t_0$ is the most appropriate point if we do not have the information about future channel status. Similar result can be straightforwardly extended for the case when $t\in[0,\frac{L_0}{v_0}]$.


%

\ifCLASSOPTIONcaptionsoff
  \newpage
\fi


\begin{thebibliography}{1}
\small{

\bibitem{Kurs_1}
A. Kurs, A. Karalis, R. Moffatt, et al., \textquotedblleft Wireless power transfer via strongly coupled magnetic resonances," \emph{Science}, vol. 317, no. 5834, pp. 83--86, Jul. 2007.

\bibitem{Grover_2}
P. Grover, and A. Sahai, \textquotedblleft Shannon meets Tesla: wireless information and power transfer," \emph{in Proc. 2010 IEEE International Symposium on Information Theory (ISIT)}, Austin, USA, 2010, pp. 2363--2367.

\bibitem{Liu_3}
L. Liu, R. Zhang, and K. C. Chua, \textquotedblleft Wireless information and power transfer: a dynamic power splitting approach," \emph{IEEE Trans. Commun.}, vol. 61, no. 9, pp. 3990--4001, Sep. 2013.

\bibitem{Fan_4}
K. Xiong, P. Y. Fan, C. Zhang, et al., "Wireless information and energy transfer for two-hop non-regenerative MIMO-OFDM relay networks," \emph{IEEE J. Sel. Areas Commun.}, vol. 33, no. 9, pp. 1595--1611, Aug. 2015.

\bibitem{Chen_5}
X. Chen, C. Yuen, and Z. Zhang,  \textquotedblleft Wireless energy and information transfer tradeoff for limited-feedback multiantenna systems with energy beamforming," \emph{IEEE Trans. Veh. Technol.}, vol. 63, no. 1, pp. 407--412, Jan. 2014.

\bibitem{Li_6}
T. Li, P. Y. Fan, and K. B. Letaief, \textquotedblleft Data acquisition with RF-based energy harvesting sensor: from information theory to green system." \emph{in Proc. 2014 IEEE Globecom}, Austin, USA, 2014, pp. 3972--3977.

\bibitem{Ju_7}
H. Ju, and R. Zhang,  \textquotedblleft Throughput maximization in wireless powered communication networks," \emph{IEEE Trans. Wirel. Commun.}, vol. 13, no. 1, pp. 418--428, Jan. 2014.

\bibitem{Shi_8}
Y. Shi, L. Xie, Y. T. Hou, and H. D. Sherali, \textquotedblleft On renewable sensor networks with wireless energy transfer," \emph{in Proc. 2011 IEEE INFOCOM}, 2011, pp. 1350--1358.

\bibitem{He_9}
S. B. He, J. M. Chen, Y. X. Sun, et al.,  \textquotedblleft On optimal information capture by energy-constrained mobile sensors," \emph{IEEE Trans. Veh. Technol.}, vol. 59, no. 5, pp. 2472--2484, Jun. 2010.

\bibitem{Bose_10}
I. Bose, and S. P. Yan, \textquotedblleft Mobile platform for the green potential of RFID projects: a case-based analysis," \emph{IT Professional}, vol. 13, no. 1, pp. 41--47, Jan. 2011.

\bibitem{Liu_11}
V. Liu, A. Parks, V. Talla, et al., \textquotedblleft Ambient backscatter: wireless communication out of thin air," \emph{in Proc. 2013 ACM SIGCOMM}, 2013, pp. 39--50.

\bibitem{Yang_12}
Y. Q. Yang, and A. E. Fathy,  \textquotedblleft Development and implementation of a real-time see-through-wall radar system based on FPGA," \emph{IEEE Trans. Geosci. Remote Sens.}, vol. 47, no. 5, pp. 1270--1280, May. 2009.

\bibitem{Sharma_13}
V. Sharma, U. Mukherji, V. Joseph and S. Gupta,  \textquotedblleft Optimal energy management policies for energy harvesting sensor nodes," \emph{IEEE Trans. Wirel. Commun.}, vol. 9, no. 4, pp. 1326--1336, Apr. 2010.

\bibitem{Yang_14}
J. Yang, O. Ozel, and S. Ulukus,  \textquotedblleft Broadcasting with an energy harvesting rechargeable transmitter," \emph{IEEE Trans. Wirel. Commun.}, vol. 11, no. 2, pp. 571--583, Feb. 2012.

\bibitem{Ding_15}
Z. G. Ding, S. M. Perlaza, I. Esnaola, and H. Vincent Poor, \textquotedblleft Power allocation strategies in energy harvesting wireless cooperative networks," \emph{IEEE Trans. Wirel. Commun.}, vol. 13, no. 2, pp. 846--860, Feb. 2014.

\bibitem{Ho_16}
C. K. Ho, and R. Zhang, \textquotedblleft Optimal energy allocation for wireless communications with energy harvesting constraints," \emph{IEEE Trans. Signal Process.}, vol. 60, no. 9, pp. 4808--4818, Sep. 2012.

\bibitem{Tutuncuoglu_17}
K. Tutuncuoglu, and A. Yener,  \textquotedblleft Optimum transmission policies for battery limited energy harvesting nodes," \emph{IEEE Trans. Wirel. Commun.}, vol. 11, no. 3, pp. 1180--1189, Mar. 2012.

\bibitem{Ozel_18}
O. Ozel, K. Tutuncuoglu, J. Yang, et al., \textquotedblleft Transmission with energy harvesting nodes in fading wireless channels: Optimal policies," \emph{IEEE J. Sel. Areas Commun.}, vol. 29, no. 8, pp. 1732--1743, Sep. 2011.

\bibitem{Huang_19}
K. Huang, and E. G. Larsson, \textquotedblleft Simultaneous information and power transfer for broadband wireless systems," \emph{IEEE Trans. Signal Process.}, vol. 61, no. 23, pp. 5972--5986, Dec. 2013.

\bibitem{Tse_20}
D. Tse, and P. Viswanath, \textquotedblleft Fundamentals of wireless communication," \emph{Cambridge University Press,} 2005.

\bibitem{Zhou_21}
X. Zhou, R. Zhang, and C. K. Ho, \textquotedblleft Wireless information and power transfer: architecture design and rate-energy tradeoff," \emph{IEEE Trans. Commun.}, vol. 61, no. 11, pp.4757--4767, Nov. 2013.

\bibitem{Rajesh_22}
R. Rajesh, V. Sharma, et al., \textquotedblleft Capacity of gaussian channels with energy harvesting and processing cost," \emph{IEEE Trans. Inf. Theory}, vol. 60, no. 5, pp.2563--2575, May. 2014.

\bibitem{Goldsmith_23}
A. Goldsmith, and P. Varaiya, \textquotedblleft Capacity of fading channel with channel side information." \emph{IEEE Trans. Inf. Theory}, vol. 43, no. 6, pp. 1986--1992, 1997.

\bibitem{Luo_24}
S. Luo, R. Zhang, and T. J. Lim, \textquotedblleft Optimal save-then-transmit protocol for energy harvesting wireless transmitters," \emph{IEEE Trans. Wirel. Commun.}, vol. 12, no. 3, pp. 1196--1207, Mar. 2013.

\bibitem{Ozel_25}
O. Ozel, and S. Ulukus, \textquotedblleft Achieving AWGN capacity under stochastic energy harvesting," \emph{IEEE Trans. Inf. Theory}, vol. 58, no. 10, pp.6471--6483, Oct. 2012.

\bibitem{Zhou_26}
X. Zhou, C. K. Ho, and R. Zhang, \textquotedblleft Wireless power meets energy harvesting: a joint energy allocation approach," \emph{in Proc. 2014 IEEE GlobalSIP}, Atlanta, USA, 2014, pp. 198--202.

\bibitem{Li_27}
T. Li, P. Y. Fan, and K. B. Letaief, \textquotedblleft Energy harvesting sensor networks with a mobile control center: optimal transmission policy." \emph{in Proc. 2015 IEEE ICC}, London, UK, 2015, pp. 3528--3533.

\bibitem{Boyd_28}
S. Boyd, and L. Vandenberghe, Convex optimization, \emph{Cambridge university press}, 2004.

\bibitem{Zhang_29}
J. Xu, and R. Zhang, \textquotedblleft Energy beamforming with one-bit feedback," \emph{IEEE Trans. Signal Process.}, vol. 62, no. 20, pp. 5370--5381, Oct. 2014.

\bibitem{Zeng_30}
Y. Zeng, and R. Zhang, \textquotedblleft Optimized training design for wireless energy transfer," \emph{IEEE Trans. Commun.}, vol. 63, no. 2, pp. 536--550, Feb. 2015.

}
\end{thebibliography}
\end{document}